\documentclass[10pt,a4paper,number,sort&compress]{elsarticle}
\usepackage{amssymb,amsmath,epsfig,color}
\usepackage{pgf,pgfarrows}
\usepackage[
  pdftitle={Conformal superspace sigma-models},
  pdfauthor={Vladimir Mitev, Thomas Quella and Volker Schomerus},
  pdfkeywords={Conformal Field Theory, Supergeometry, Harmonic Analysis},
  pdfdisplaydoctitle=true]{hyperref}

\DeclareMathOperator{\End}{\text{End}}
\DeclareMathOperator{\Hom}{\text{Hom}}
\DeclareMathOperator{\Ind}{\text{Ind}}
\DeclareMathOperator{\Inv}{\text{Inv}}
\DeclareMathOperator{\tr}{\text{tr}}
\DeclareMathOperator{\str}{\text{str}}
\DeclareMathOperator{\Ad}{{\text{Ad}}}
\DeclareMathOperator{\ad}{{\text{ad}}}
\DeclareMathOperator{\id}{{\text{id}}}

\newcommand{\Real}{\mathbb{R}}
\newcommand{\Complex}{\mathbb{C}}

\newcommand{\Integer}{\mathbb{Z}}

\newcommand{\bartial}{{\bar{\partial}}}

\newcommand\Rep{\text{Rep}}

\newcommand{\g}{\mathfrak{g}}

\newcommand{\cA}{\mathcal{A}}
\newcommand{\cB}{\mathcal{B}}

\newcommand{\cF}{\mathcal{F}}  
  
\newcommand{\cH}{\mathcal{H}}  
\newcommand{\cI}{\mathcal{I}}  

\newcommand{\cK}{\mathcal{K}}
\newcommand{\cL}{\mathcal{L}}
\newcommand{\cM}{\mathcal{M}}

\newcommand{\cP}{\mathcal{P}}

\newcommand{\cS}{\mathcal{S}}

\newcommand{\bz}{{\bar{z}}}
\newcommand{\bw}{{\bar{w}}}

\newcommand{\bh}{{\bar{h}}}
\newcommand{\bJ}{{\bar{J}}}
\newcommand{\Projective}{\mathbb{P}}

\newcommand{\Typ}{\text{Typ}}
\newcommand{\ATyp}{\text{Atyp}}

\definecolor{LightRed}{rgb}{1.0,0.5,0.5}
\definecolor{DarkGreen}{rgb}{0.0,0.90,0.0}
\definecolor{LemonChiffon}{rgb}{1.0,0.9,0.8}
\definecolor{LightGreen}{rgb}{.798,1.0,0.4}
\definecolor{Orange}{rgb}{1,0.6,0.0}
\definecolor{grey}{rgb}{.8,0.9,0.9}
\definecolor{pink}{rgb}{1,0.753,0.796}

\newcommand{\red}{\color{red}}

\newcommand{\violet}{\color{violet}}

\title{Conformal superspace $\sigma$-models\tnoteref{t1}}

\author[desy]{Vladimir Mitev\fnref{fn1}}
\ead{Vladimir.Mitev@desy.de}
  
\author[kdv]{Thomas Quella\corref{cor1}\fnref{fn2}}
\ead{T.Quella@uva.nl}

\author[desy]{Volker Schomerus}
\ead{Volker.Schomerus@desy.de}

\address[desy]{DESY Hamburg, Theory Group, Notkestra\ss{}e 85, D--22607 Hamburg, Germany}
\address[kdv]{Korteweg de Vries Instituut voor Wiskunde, Universiteit van Amsterdam,\\
PO Box 94248, 1090 GE Amsterdam, The Netherlands}

\cortext[cor1]{Corresponding author. Phone: +49-221-470-7420
Fax: +49-221-470-5159, Email: Thomas.Quella@uni-koeln.de.}

\tnotetext[t1]{Talk given by
    Thomas Quella at the Lorentz Center Workshop ``The Interface of
    Integrability and Quantization'' (Leiden, 12.-16.4.2010) on the
    occasion of the 60th birthday of G.F.\ Helminck.}

\fntext[fn1]{Present address: Institut f\"ur Mathematik,
  Humboldt-Universit\"at zu Berlin, Rudower Chaussee 25, D-12489
  Berlin, Germany.}
\fntext[fn2]{Present address: Institut f\"ur Theoretische Physik,
  Universit\"at zu K\"oln, Z\"ulpicher Stra\ss{}e 77, D-50937 Cologne,
  Germany.}

\date{}

\begin{document}

\begin{abstract}
  We review recent developments in the context of two-dimensional
  conformally invariant $\sigma$-models. These quantum field theories
  play a prominent role in the covariant superstring quantization in
  flux backgrounds and in the analysis of disordered systems.
 
  We present supergroup WZW models as
  primary examples of logarithmic conformal field theories, whose
  structure is almost entirely determined by the underlying
  supergeometry. In particular, we discuss the harmonic analysis
  on supergroups and supercosets and point out the subtleties of Lie
  superalgebra representation theory that are responsible for the
  emergence of logarithmic representations. Furthermore,
  special types of marginal deformations of supergroup WZW models are
  studied which only exist if the Killing form is vanishing. We show
  how exact expressions for anomalous dimensions of boundary fields
  can be derived using quasi-abelian perturbation theory. Finally,
  the knowledge of the exact spectrum is used to motivate a duality
  between the $\text{OSP}(4|2)$ symmetric Gross-Neveu model and the
  $\text{S}^{3|2}$ supersphere $\sigma$-model.
  \\[3mm]
  [MSC numbers: 17Bxx, 81T40]
\end{abstract}


  
\maketitle

\section{Introduction}

  The development of exactly solvable two-dimensional quantum field
  theories provides one of the nicest and most fruitful examples of
  a co-evolution of physics and mathematics. Due to the fundamental
  importance of such models in string theory, a promising candidate
  for a theory of quantum gravity, the subject naturally
  connects geometrical, topological and algebraic questions. On the
  algebraic side it is naturally related to quantum groups as well as
  infinite dimensional Lie algebras such as affine Kac-Moody algebras
  and the Virasoro algebra. On the other hand, simple algebraic
  manipulations such as applying an automorphism of a chiral vertex
  algebra, are intimately related to deep geometric phenomena such as
  mirror symmetry between Calabi-Yau manifolds.
  
  In this note we discuss the mathematical structures that arise in a
  special class of two-dimensional quantum field theories, namely
  conformally invariant superspace $\sigma$-models. The sole inclusion
  of the word ``super'' leads to a number of subtleties and features
  which do not have counterparts in comparable bosonic models. One of
  the issues addressed in this article is the harmonic analysis on
  supergroups and supercosets which determines the $\sigma$-model
  spectrum at large
  volume. In particular, we exhibit the existence of logarithmic
  representations as an immediate consequence of the underlying
  supergeometry. We then discuss two different types of deformations
  of supergroup WZW models, which require the supergroup to have
  a vanishing Killing form. In both cases we are able to determine
  certain open string partition functions exactly for all values of
  the coupling. We finally use these results to argue for a duality
  between the $\text{OSP}(4|2)$ Gross-Neveu model and the $\text{S}^{3|2}$
  supersphere $\sigma$-model.

  The types of $\sigma$-models discussed in this note are an essential
  ingredient in the covariant quantization of superstrings in flux
  backgrounds, specifically in various types of Anti-de\,Sitter
  spaces, see table \ref{tab:Strings}. Further applications exist in
  condensed matter theory (see table \ref{tab:CondMat}), where
  Efetov's supersymmetry trick allows us to express physical
  observables in disordered systems (such as quantum Hall systems) in
  terms of correlation functions in a superspace $\sigma$-model
  \cite{Efetov1983:MR708812} (see also
  \cite{Parisi:1979ka,Parisi:1982ud}). Also the universality classes
  of certain loop ensembles can be described by this class of models
  \cite{Read:2001pz,Candu:2009pj}.

\begin{table}
\begin{center}
  \begin{tabular}{c|c|c|c|c}
    Minkowski & $\text{AdS}_5\times\text{S}^5$ &
    $\text{AdS}_4\times\Complex\Projective^3$ & $\text{AdS}_3\times
    S^3$ & $\text{AdS}_3\times\text{S}^3\times\text{S}^3$\\\hline&&&\\[-3mm]
     $\frac{\text{super-Poincar\'e}}{\text{Lorentz}}$&$\frac{\text{PSU}(2,2|4)}{\text{SO}(1,4)\times
       \text{SO}(5)}$ & $\frac{\text{OSP}(6|2,2)}{\text{U}(3)\times
       \text{SO}(1,3)}$ & $\text{PSU}(1,1|2)$ & $\text{D}(2,1;\alpha)$
  \end{tabular}
  \caption{\label{tab:Strings}Supercosets and their applications in
    string theory. The supercosets and supergroups in the lower row
    describe a supersymmetrized version of the geometries in the first
    line.}
\end{center}
\end{table}

\begin{table}
\begin{center}
  \begin{tabular}{c|c|c}
    IQHE & Dilute polymers (SAW)& Dense polymers\\\hline&&\\[-3mm]
     {\tiny(non-conformal)}&$\text{S}^{2S+1|2S}$&$\Complex\Projective^{S-1|S}$\\[2mm]
     $\frac{\text{U}(1,1|2)}{\text{U}(1|1)\times\text{U}(1|1)}$&$\frac{\text{OSP}(2S+2|2S)}{\text{OSP}(2S+1|2S)}$&$\frac{\text{U}(S|S)}{\text{U}(1)\times\text{U}(S-1|S)}$
  \end{tabular}
  \caption{\label{tab:CondMat}Supercosets and their applications in condensed matter
    theory and statistical physics.}
\end{center}
\end{table}

\section{Non-linear \texorpdfstring{$\sigma$}{sigma}-models on superspaces}

  In this section we provide a brief introduction to superspace
  $\sigma$-models. We then discuss issues related to conformal
  invariance and discuss the moduli space of such theories. Circle
  theories are used as an illustrative example where exact
  results can be obtained.

\subsection{Superspace  \texorpdfstring{$\sigma$}{sigma}-models}

  Non-linear $\sigma$-models are quantum field theories describing the
  embedding of a two-dimensional surface $\Sigma$, the world-sheet,
  into some (pseudo-)Riemannian (super)manifold $\cM$. The latter may
  be equipped with some additional structure besides the
  metric. Examples include vector bundles and gerbes.
  In the case $\Sigma$ is a cylinder, the model describes the
  propagation of a closed string in the ``universe'' $\cM$, in the case
  of a strip the propagation of an open string. More complicated
  world-sheets with multiple holes and boundary components have to be
  used to describe the interaction between different strings.

  The dynamics of the world-sheet is governed by an action functional
  $\cS$ in which the metric and the other structures of the manifold
  $\cM$ enter as parameters. In the simplest case, the functional
  $\cS$ reads
\begin{align}
  \cS[X]
  \ =\ \frac{1}{2\pi}\int\!d^2z
       \Bigl[G_{\mu\nu}(X)+B_{\mu\nu}(X)\Bigr]\partial X^\mu\bartial X^\nu\ \ .
\end{align}
  It assigns a real number to each embedding $X:\Sigma\to\cM$. The
  model can be regarded as a two-dimensional quantum field theory with
  background fields $G(X)$ (the metric) and $B(X)$ (a two-form)
  playing the role of coupling constants.

  Besides their physical importance for the description of strings,
  non-linear $\sigma$-models are also interesting from a mathematical
  point of view since they probe the symmetries and the topology of
  $\cM$. Indeed, the isometries of the space $\cM$ reflect themselves
  as internal symmetries of the quantum field theory. Moreover, both
  open and closed strings as well as D-branes can wind around
  non-trivial cycles in $\cM$. The nature and size of these cycles are
  encoded in the spectrum of string excitations.

  While the geometric interpretation of the $\sigma$-model is clear at
  large scales, it becomes less obvious on smaller scales where
  quantum effects become important. In a sense, non-linear
  $\sigma$-models provide a way of defining a notion of ``quantum
  geometry''. Some of the surprises of quantum geometry are sketched
  below, e.g.\ the fact that a small circle cannot be distinguished
  from a small three-sphere if the sizes are chosen appropriately.

\subsection{\label{sc:ConformalInvariance}Conformal invariance}

\begin{figure}
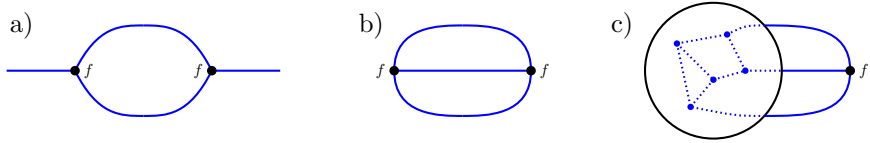

\begin{center}
\begin{pgfpicture}{0cm}{.2cm}{16cm}{1.4cm}
  \pgfputat{\pgfxy(0.2,1.2)}{\pgfbox[center,center]{a)}}
  \pgfputat{\pgfxy(4.8,1.2)}{\pgfbox[center,center]{b)}}
  \pgfputat{\pgfxy(8.1,1.2)}{\pgfbox[center,center]{c)}}
\begin{pgfmagnify}{0.6}{0.6}
  \pgfsetlinewidth{1.3pt}
  \color{blue}
  \pgfline{\pgfxy(0,1)}{\pgfxy(1.5,1)}
  \pgfmoveto{\pgfxy(1.5,1)}
  \pgfcurveto{\pgfxy(2,2)}{\pgfxy(2.5,2)}{\pgfxy(3,2)}
  \pgfstroke
  \pgfmoveto{\pgfxy(1.5,1)}
  \pgfcurveto{\pgfxy(2,0)}{\pgfxy(2.5,0)}{\pgfxy(3,0)}
  \pgfstroke
  \color{blue}
  \pgfline{\pgfxy(4.5,1)}{\pgfxy(6,1)}
  \pgfmoveto{\pgfxy(4.5,1)}
  \pgfcurveto{\pgfxy(4,2)}{\pgfxy(3.5,2)}{\pgfxy(3,2)}
  \pgfstroke
  \pgfmoveto{\pgfxy(4.5,1)}
  \pgfcurveto{\pgfxy(4,0)}{\pgfxy(3.5,0)}{\pgfxy(3,0)}
  \pgfstroke
  \color{black}
  \pgfcircle[fill]{\pgfxy(1.5,1)}{3pt}
  \pgfcircle[fill]{\pgfxy(4.5,1)}{3pt}
  \pgfputat{\pgfxy(1.8,1)}{\pgfbox[center,center]{$f$}}
  \pgfputat{\pgfxy(4.2,1)}{\pgfbox[center,center]{$f$}}
\begin{pgftranslate}{\pgfpoint{7cm}{-.5cm}}
  \pgfsetlinewidth{1.3pt}
  \color{blue}
  \pgfline{\pgfxy(1.5,1.5)}{\pgfxy(4.5,1.5)}
  \pgfmoveto{\pgfxy(1.5,1.5)}
  \pgfcurveto{\pgfxy(1.5,2.5)}{\pgfxy(2.5,2.5)}{\pgfxy(3,2.5)}
  \pgfstroke
  \pgfmoveto{\pgfxy(1.5,1.5)}
  \pgfcurveto{\pgfxy(1.5,0.5)}{\pgfxy(2.5,0.5)}{\pgfxy(3,0.5)}
  \pgfstroke
  \color{blue}
  \pgfmoveto{\pgfxy(4.5,1.5)}
  \pgfcurveto{\pgfxy(4.5,2.5)}{\pgfxy(3.5,2.5)}{\pgfxy(3,2.5)}
  \pgfstroke
  \pgfmoveto{\pgfxy(4.5,1.5)}
  \pgfcurveto{\pgfxy(4.5,0.5)}{\pgfxy(3.5,0.5)}{\pgfxy(3,0.5)}
  \pgfstroke
  \color{black}
  \pgfcircle[fill]{\pgfxy(1.5,1.5)}{3pt}
  \pgfcircle[fill]{\pgfxy(4.5,1.5)}{3pt}
  \pgfputat{\pgfxy(1.2,1.5)}{\pgfbox[center,center]{$f$}}
  \pgfputat{\pgfxy(4.8,1.5)}{\pgfbox[center,center]{$f$}}
\end{pgftranslate}
\begin{pgftranslate}{\pgfpoint{14cm}{-.5cm}}
  \pgfsetlinewidth{1.3pt}
  \color{blue}
  \pgfline{\pgfxy(3,1.5)}{\pgfxy(4.5,1.5)}
  \color{blue}
  \pgfmoveto{\pgfxy(4.5,1.5)}
  \pgfcurveto{\pgfxy(4.5,2.5)}{\pgfxy(3.5,2.5)}{\pgfxy(2.63,2.5)}
  \pgfstroke
  \pgfmoveto{\pgfxy(4.5,1.5)}
  \pgfcurveto{\pgfxy(4.5,0.5)}{\pgfxy(3.5,0.5)}{\pgfxy(2.63,0.5)}
  \pgfstroke
  \color{black}
  \pgfcircle[fill]{\pgfxy(4.5,1.5)}{3pt}
  \pgfputat{\pgfxy(4.8,1.5)}{\pgfbox[center,center]{$f$}}
  \pgfstroke
  \color{black}
  \pgfcircle{\pgfxy(1.5,1.5)}{1.5cm}
  \pgfstroke
  \color{blue}
  \pgfsetdash{{1pt}{2pt}}{0pt}  
  \pgfmoveto{\pgfxy(2.63,0.5)}
  \pgfcurveto{\pgfxy(2,0.5)}{\pgfxy(2,0.5)}{\pgfxy(1,0.7)}
  \pgfmoveto{\pgfxy(2.63,2.5)}
  \pgfcurveto{\pgfxy(2.2,2.5)}{\pgfxy(2.2,2.5)}{\pgfxy(1.8,2.3)}
  \pgfline{\pgfxy(2.2,1.5)}{\pgfxy(1.8,2.3)}
  \pgfline{\pgfxy(2.2,1.5)}{\pgfxy(3,1.5)}
  \pgfline{\pgfxy(0.7,2.1)}{\pgfxy(1.8,2.3)}
  \pgfline{\pgfxy(0.7,2.1)}{\pgfxy(1.5,1.3)}
  \pgfline{\pgfxy(0.7,2.1)}{\pgfxy(1,0.7)}
  \pgfline{\pgfxy(1.5,1.3)}{\pgfxy(1,0.7)}
  \pgfline{\pgfxy(1.5,1.3)}{\pgfxy(2.2,1.5)}
  \pgfstroke
  \pgfcircle[fill]{\pgfxy(1.5,1.3)}{2pt}
  \pgfcircle[fill]{\pgfxy(.7,2.1)}{2pt}
  \pgfcircle[fill]{\pgfxy(1,0.7)}{2pt}
  \pgfcircle[fill]{\pgfxy(1.8,2.3)}{2pt}
  \pgfcircle[fill]{\pgfxy(2.2,1.5)}{2pt}
  \pgfstroke
\end{pgftranslate}
\end{pgfmagnify}
\end{pgfpicture}
\end{center}
  \caption{\label{fig:KillingForm}a) The Killing form.
    b) \& c) Vanishing contributions to the $\beta$-function.}
\end{figure}

  In general, non-linear $\sigma$-models on arbitrary spaces $\cM$
  are not solvable. They are complicated interacting quantum field
  theories whose couplings obey intricate perturbative renormalization
  group equations. For this reason we wish to restrict our attention
  to a special class of $\sigma$-models whose extended symmetry
  considerably reduces the complexity of the problem. More
  precisely, we wish to consider two specific classes of
  conformally invariant $\sigma$-models. In two dimensions, the
  conformal invariance implies the existence of an infinite
  dimensional spectrum generating algebra, diminishing
  the actual number of independent degrees of freedom.

\begin{figure}
\begin{center}
\begin{pgfpicture}{0cm}{-.2cm}{10cm}{2.4cm}
  \pgfputat{\pgfxy(7,1.2)}{\pgfbox[center,center]{\pgfimage[height=2cm]{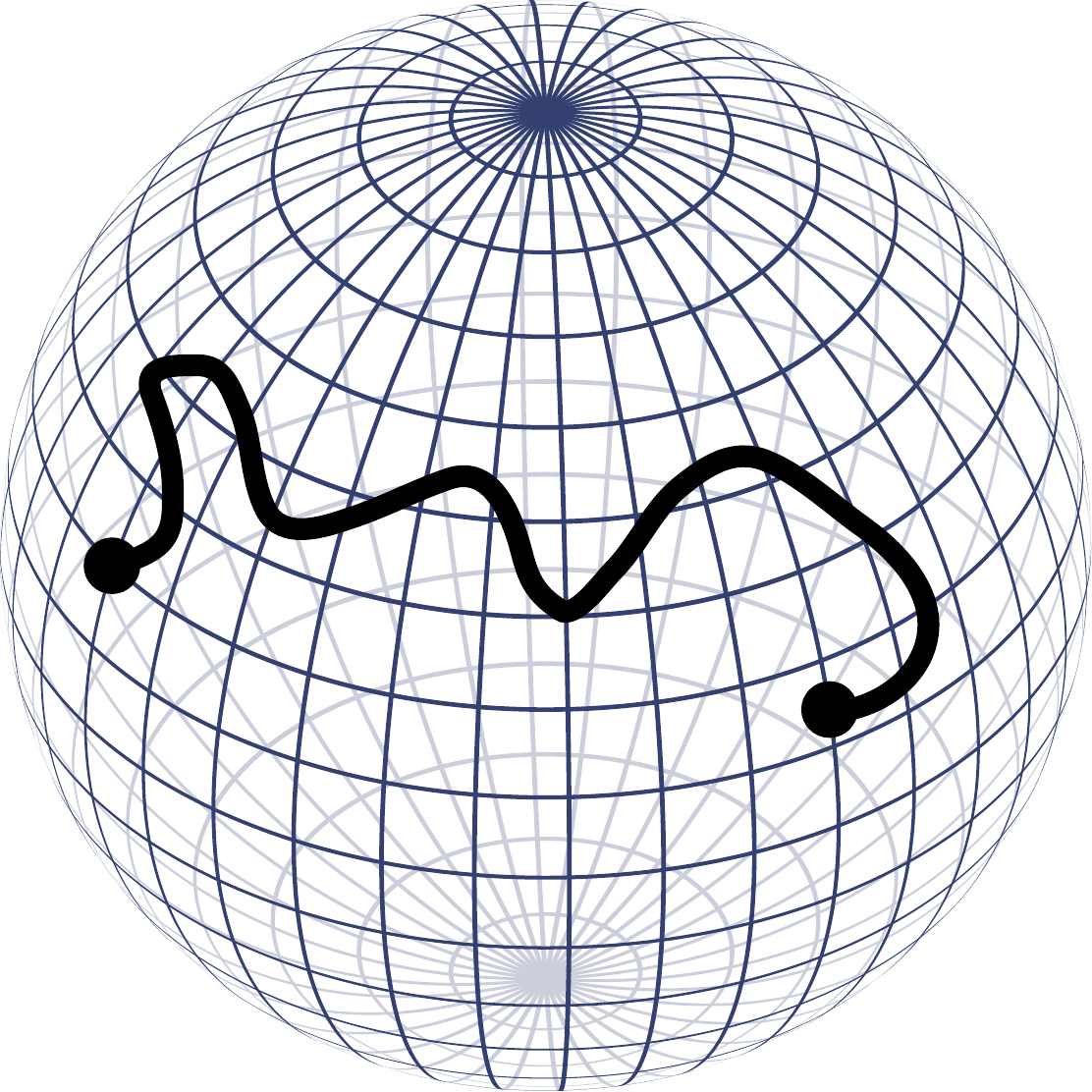}}}  
  \pgfputat{\pgfxy(4,1.2)}{\pgfbox[center,center]{\pgfimage[height=1cm]{WireSphereWithString}}}  
  \pgfputat{\pgfxy(1.5,1.2)}{\pgfbox[center,center]{\pgfimage[height=.16cm]{WireSphereWithString}}}  
  \pgfsetendarrow{\pgfarrowtriangle{3pt}}
  \pgfsetlinewidth{1.2pt}
  \pgfline{\pgfxy(1,0)}{\pgfxy(9,0)}
  \pgfputat{\pgfxy(9.8,0)}{\pgfbox[center,center]{Radii}}
  \pgfputat{\pgfxy(2.9,-0.5)}{\pgfbox[center,center]{Quantum regime}}
  \pgfputat{\pgfxy(7,-0.5)}{\pgfbox[center,center]{Classical regime}}
  \pgfclearendarrow
  {\pgfcircle[fill]{\pgfxy(1,0)}{2pt}}
  \pgfsetlinewidth{1.2pt}
  \pgfstroke
  \pgfsetdash{{1pt}{2pt}}{0pt}  
\end{pgfpicture}
\end{center}
  \caption{\label{fig:Moduli}The moduli space of generic supercoset
    $\sigma$-models.}
\end{figure}

  The two classes of target spaces that are considered in this note
  are supergroups
  $\cM=\text{G}$ and supercosets $\cM=\text{G/H}$.\footnote{Note that
    a supergroup $\text{G}$ can be realized as a diagonal supercoset
    $\text{G}\times\text{G}/\text{G}$. For reasons that become clear
    below, it is nevertheless useful to distinguish these two cases.}
  In the latter case, the identification is given by $g\sim gh$, with
  $g\in\text{G}$ and $h\in\text{H}$. The spaces $\text{G}$ and
  $\text{G/H}$ have isometries $\text{G}\times\text{G}$ and
  $\text{G}$, respectively, which act as left (and right in the first
  case) multiplication. In order to guarantee conformal invariance
  some extra conditions on the metric and on the $B$-field have to be
  met. Since we impose $\text{G}$-invariance we only have one
  parameter at our disposal for the metric in case $\text{G}$ is a
  simple supergroup. Indeed, the metric is uniquely determined up to a
  scalar in that case. In most of the examples conformal invariance
  uniquely determines $B$ in terms of the metric. In these cases the
  volume of $\text{G}$ will hence be the only modulus,
  cf.~Fig.~\ref{fig:Moduli}. Exceptions are the deformations of
  supergroup WZW models discussed in section \ref{sc:Deformations}
  below where two moduli are available.

  In the case of supercosets, it is custom to assume that $\text{H}$ is the
  fixed point set under some finite order automorphism of $\text{G}$. To
  first order in perturbation theory, the associated non-linear
  $\sigma$-models are conformally invariant
  if the Killing form of $\text{G}$ is vanishing \cite{Kagan:2005wt}. After
  fixing a basis $T^a$ of the Lie superalgebra $\g$ underlying $\text{G}$,
  this statement is equivalent to
\begin{align}
  \label{eq:KillingForm}
  K^{ab}
  \ =\ \str(\ad_{T^a}\circ\ad_{T^b})
  \ =\ -(-1)^{\text{deg}(d)}{f^{ac}}_d{f^{bd}}_c
  \ =\ 0\ \ .
\end{align}
  For many supercosets relevant in physical applications (see tables
  \ref{tab:Strings} and \ref{tab:CondMat}) this
  condition is even sufficient to all orders (cf.~the tables in
  \cite{Candu:2010yg}). While for a purely bosonic Lie group
  eq.~\eqref{eq:KillingForm} can never be satisfied if $\text{G}$ is simple,
  it is well possible for certain families of Lie supergroups. These
  series are $\text{PSL}(N|N)$, $\text{OSP}(2S+S|2S)$ and
  $\text{D}(2,1;\alpha)$.\footnote{There are further series which, however, are
    not interesting from a physical point of view since they do not
    admit a non-degenerate metric.} The interesting physical
  applications of these supergroups are displayed in the tables
  \ref{tab:Strings} and \ref{tab:CondMat}.

  The condition \eqref{eq:KillingForm} for conformal invariance has a
  nice diagrammatic interpretation, see part a) of figure
  \ref{fig:KillingForm} for an illustration of the Killing form.
  Intuitively, it is now clear why the $\beta$-function does not
  receive corrections. As an invariant with respect to the global
  $\text{G}$-symmetry it is defined as a sum over Feynman diagrams  without
  external legs. All these diagrams are constructed from straight
  lines and trivalent vertices, corresponding to the metric
  and the structure constants ${f^{ab}}_c$ on $\g$, respectively.
  Considering a
  potential contribution involving at least one vertex, the latter has
  to be connected to the rest of the diagram through three legs. The
  contribution in part b) of figure \ref{fig:KillingForm} clearly
  vanishes. Let us therefore consider a general diagram as in part c) of
  the same figure. For the cases at hand, there is just one invariant
  three-tensor, and hence the ``blob'' will be proportional to the
  structure constants $f$. The diagram is then essentially equivalent
  to the one
  in part b) and hence vanishes, provided that
  eq.~\eqref{eq:KillingForm} is satisfied
  \cite{Bershadsky:1999hk}. The remaining diagrams without trivalent
  vertices can also be found in an abelian theory for which the
  $\beta$-function is known to be zero.

\subsection{Spectra and partition functions}

  All information about a conformal field theory is contained in its
  correlation functions. The quantity we are aiming for in this
  note is a boundary partition function, or in other words the vacuum
  correlation function on an annulus. This kind of partition function
  encodes the spectrum of open strings starting and ending on a given
  D-brane. The energies of the excitations, the conformal dimensions
  $h$ of the states, will depend on the moduli of $\cM$. In
  particular, they will change if the volume of the manifold is
  increased or decreased. Below the moduli will be collectively
  referred to as the ``radius'' $R$.

  The D-branes we will consider in this note preserve a global
  isometry $\text{G}$ in addition to the conformal symmetry (represented by
  the Virasoro algebra $\text{Vir}$). The string excitations can
  therefore be organized by their transformation behavior with respect
  to the total symmetry $\g\oplus\text{Vir}$. The partition function
  can thus be written as
\begin{equation}
  \label{eq:PF}
  Z(q,z|R)
  \ =\ \tr_{\cH}\Bigl[z^{H}\,q^{L_0-\frac{c}{24}}\Bigr]
  \ =\
  \sum_{\Lambda}\underbrace{\psi_\Lambda(q,R)}_{\text{Dynamics}}\,\underbrace{\chi_\Lambda(z)}_{\text{Symmetry}}\
  \ ,
\end{equation}
  where  $\chi_\Lambda(z)$ are characters of $\g$.
  In this expression, $L_0$ and $c$ refer to the energy operator and
  the central charge of the Virasoro algebra. The second insertion
  $z^H=\prod z_i^{H_i}$ keeps track of the eigenvalues of the Cartan
  generators $H_i$ of $\g$. On the left hand side and the right hand
  side we made explicit that the partition function depends on the
  moduli of $\cM$. In the middle, this is implicit since the state
  space $\cH$ will depend on $R$.

  At large values of $R$, the partition function \eqref{eq:PF} is
  entirely determined by the geometric and topological properties of
  $\cM$. In particular, the energy of string excitations can be
  obtained by performing a harmonic analysis on $\cM$. For small
  values of $R$, however, the $\sigma$-model becomes strongly coupled
  and geometry starts to loose its meaning (see the example below). In
  this regime, calculating the partition function is a highly
  non-trivial task. While the character $\chi_\Lambda(z)$ only reflect
  the symmetries of the model, the dynamical information is
  contained in the branching functions $\psi_\Lambda(q,R)$. These
  functions describe which multiplets are located at which energy
  level for a given value of the moduli $R$. An example of a spectrum
  is sketched below in figure \ref{fig:Spectrum}.

\subsection{An example: The circle  \texorpdfstring{$S^1$}{S(1)}}

  The simplest non-trivial target space is the circle $S^1$. Since the
  circle does not support a $B$-field, the non-linear $\sigma$-model
  only has one parameter $R$, the circle radius. Since moreover the
  circle is flat (i.e.\ it has no intrinsic curvature), the theory is
  described by a {\em free} boson. This theory can be solved exactly
  for all values of the radius $R$ using the underlying
  $\widehat{U}(1)$ current algebra
\begin{align}
  \label{eq:FreeOPE}
  J(z)\,J(w)
  \ =\ \frac{1}{(z-w)^2}\ \ .
\end{align}
  The partition functions of an open string with free boundary
  conditions on both ends turns out to be
\begin{align}
  \label{eq:BosonPF}
  Z(q,z|R)
  &\ =\ \tr_{\cH}\Bigl[z^{P}\,q^{L_0-\frac{c}{24}}\Bigr]
   \ =\ \frac{1}{\eta(q)}\sum_{w\in\Integer}z^w\,q^{\frac{w^2}{2R^2}}\
   \ .
\end{align}
  In this formula, $w$ denotes the eigenvalues of the quantized
  momentum quantum operator $P$ on a circle and
  the factor $\eta(q)$ keeps track of the energies of string
  oscillation modes. Note that the existence of $P$ is related to the
  $\text{U}(1)$ isometry of the circle.

\begin{figure}
\begin{center}
\begin{pgfpicture}{0cm}{-1cm}{11cm}{2.3cm}
  \pgfsetendarrow{\pgfarrowtriangle{3pt}}
  \pgfsetlinewidth{1.2pt}
  \pgfline{\pgfxy(1,0)}{\pgfxy(9,0)}
  \pgfputat{\pgfxy(9.8,0)}{\pgfbox[center,center]{Radius}}
  \pgfputat{\pgfxy(2.4,-0.5)}{\pgfbox[center,center]{Quantum regime}}
  \pgfputat{\pgfxy(7.5,-0.5)}{\pgfbox[center,center]{Classical regime}}
  \pgfclearendarrow
  {\pgfcircle[fill]{\pgfxy(1,0)}{2pt}}
  {\pgfcircle[fill]{\pgfxy(4,0)}{2pt}}
  \pgfxyline(4,1.2)(4.5,1.2)
  \pgfxyline(7,1.2)(8,1.2)
  \pgfsetendarrow{\pgfarrowto}
  \pgfclearendarrow
  \pgfputat{\pgfxy(2.2,1.2)}{\pgfbox[center,center]{$R_0^2/R$}}
  \pgfputat{\pgfxy(4.2,1.4)}{\pgfbox[center,center]{$R_0$}}
  \pgfputat{\pgfxy(7.5,1.4)}{\pgfbox[center,center]{$R$}}
  \pgfsetlinewidth{1.2pt}
  \pgfcircle{\pgfxy(7,1.2)}{1cm}
  \pgfcircle{\pgfxy(4,1.2)}{.5cm}
  \pgfcircle{\pgfxy(1.5,1.2)}{.08cm}
  \pgfcircle{\pgfxy(3,-1.3)}{.23cm}
  \pgfstroke
  \pgfsetdash{{1pt}{2pt}}{0pt}  
  \pgfxyline(1.5,1)(1.5,.1)
  \pgfxyline(4,1.2)(4,.1)
  \pgfxyline(7,1.2)(7,.1)
  \pgfsetendarrow{\pgfarrowtriangle{3pt}}
  {\pgfxyline(4,-1)(4,-.3)}
  \pgfclearendarrow
  {\pgfputat{\pgfxy(4,-1.3)}{\pgfbox[center,center]{$S^1\cong
        S^3$}}}
  {\pgfputat{\pgfxy(2.5,-1.3)}{\pgfbox[right,center]{$\widehat{\text{U}}(1)$}}}
  {\pgfputat{\pgfxy(8.2,-1.3)}{\pgfbox[right,center]{$\widehat{\text{U}}(1)$}}}
  {\pgfputat{\pgfxy(5.5,-1.3)}{\pgfbox[left,center]{$\widehat{\text{SU}}(2)_1$}}}
  \pgfputat{\pgfxy(5,-1.3)}{\pgfbox[center,center]{\pgfimage[height=.5cm]{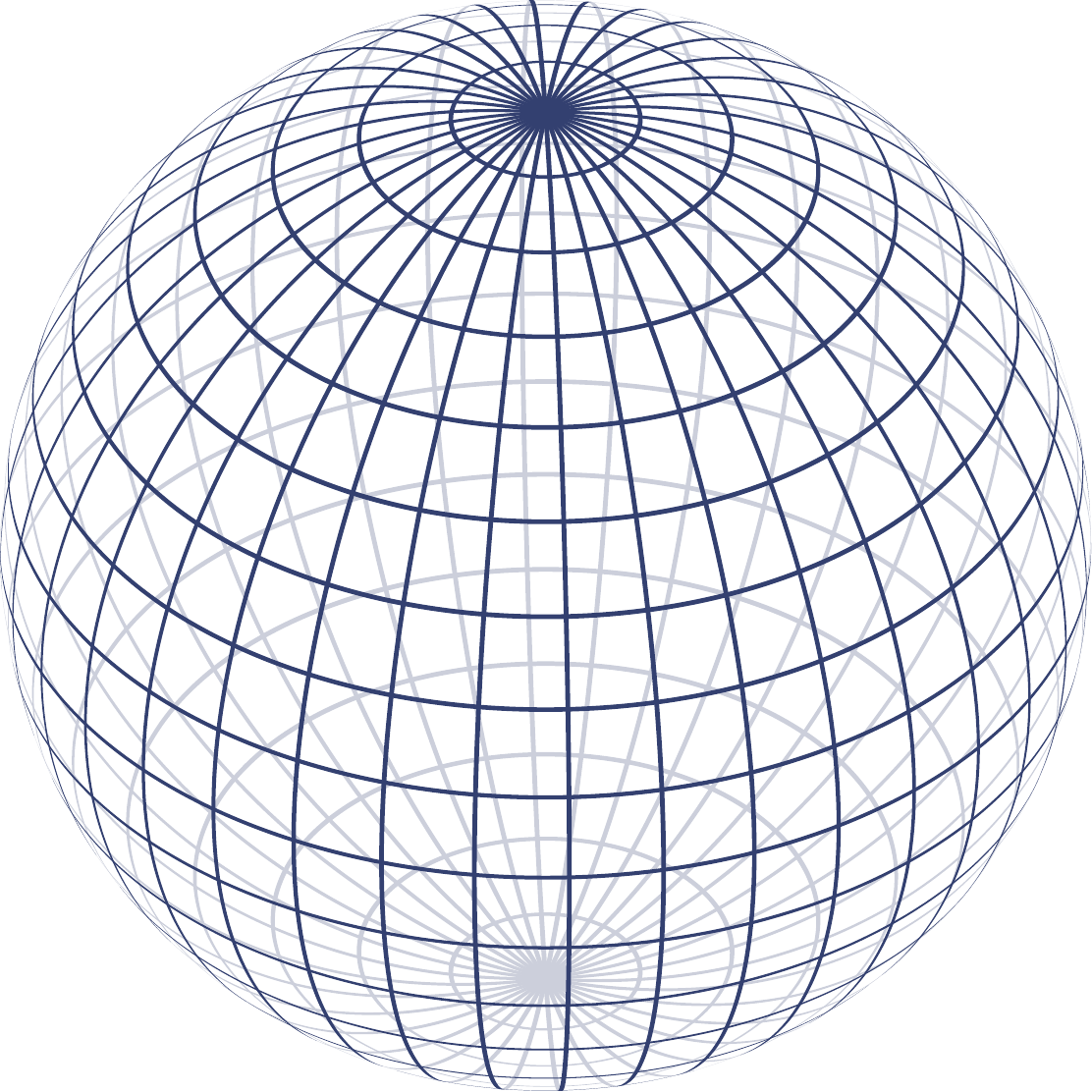}}}  
\end{pgfpicture}
\end{center}
  \caption{\label{fig:FreeBoson}The moduli space of circle theories
    and the corresponding symmetries.}
\end{figure}

  Even though the circle theory is free and therefore trivial to
  solve, it nevertheless exhibits a few surprises. The first one
  concerns an exact equivalence under the replacement
  $R\leftrightarrow R_0^2/R$ (and a simultaneous exchange of momentum
  and winding modes as well as free and fixed boundary conditions),
  also known as T-duality. In this case it implies the unexpected
  statement that the classical regime (very large $R$) is equivalent
  to the quantum regime (very small $R$). The second one concerns a
  symmetry enhancement from $\widehat{\text{U}}(1)$ to
  $\widehat{\text{SU}}(2)_1$ at the self-dual radius $R=R_0$. At this
  special point, the circle theory can be identified with a
  $\widehat{\text{SU}}(2)_1$ WZW model which in turn can be thought of
  as describing a special point in the quantum regime of
  $\sigma$-models on $S^3$. A sketch of the free boson moduli space
  can be found in figure \ref{fig:FreeBoson}. We will later use the
  analogy with the free boson to argue for the existence of a much
  more complicated equivalence between two seemingly different
  conformal field theories, one geometric, one non-geometric.

\section{Harmonic analysis on supergroups and supercosets}

  This section focuses on mathematical aspects of superspace
  $\sigma$-models. We review the representation theory of Lie
  superalgebras as well as the harmonic analysis on supergroups and
  supercosets, specifically superspheres. The results of this section
  have direct implications for the large volume spectra of superspace
  $\sigma$-models but they are also interesting from a purely
  mathematical point of view.

\subsection{Representation theory of Lie superalgebras}

  The finite dimensional simple modules of a simple Lie
  superalgebra $\g=\g_0\oplus\g_1$ fall into two classes, typical and
  atypical representations.\footnote{In physics terminology, one would
    call them non-BPS and BPS representations or long and (semi-)short
    representations, respectively.}
  The typical sector is characterized by the property that its
  simple modules are projective (in the sense of category
  theory) while for the atypical representations this is not the
  case. The atypical modules can be thought of as arising from
  degenerate limits of typical modules where the typical
  representation becomes non-semisimple.\footnote{In physics
    terminology the long multiplet splits into
    short multiplets at the BPS bound, even though this seems to
    suggest a decomposition into irreducible components
    which certainly is not the case.} Alternative characterizations in
  terms of highest weights exist \cite{Kac:1977em} (and are more
  common), but they will not be needed here.

  For the sake of completeness let us also provide some further
  details on the notion of a projective module. A $\g$-module $\cP$ is
  called projective if and only if for every surjective
  $\g$-homomorphism $f:M\twoheadrightarrow\cP$ there 
  exists a $\g$-homomorphism $h:\cP\to M$ such that $f\circ
  h=\id$. In other words, in case $M$ is a cover of $\cP$ then it
  contains $\cP$ as a direct summand and the map $f$ can be thought of
  as the projection onto $\cP$. In the present context, i.e.\ 
  restricting all considerations to the category of finite-dimensional
  $\g$-modules, projective modules also satisfy the dual property of
  being injective. One could then replace our previous definition by
  the requirement that any projective submodule $\cP$ of an arbitrary
  module $M$ always appears as a direct summand.


  Let $\Rep(\g)$ denote the set of (equivalence classes of) all
  finite dimensional simple modules. Elements from this set will be
  denoted by $\cL_\mu$, with $\mu$ running through some set of
  weights. We split the weights into typical ones and atypical ones,
  $\Typ(\g)$ and $\ATyp(\g)$. To each weight $\mu$, one can associate
  precisely one simple module $\cL_\mu$ and further indecomposable
  modules which contain $\cL_\mu$ as a simple quotient. The most
  important of these modules is the projective cover $\cP_\mu$. A
  representation $\cL_\mu$ is typical if and only if
  $\cL_\mu\cong\cP_\mu$. A further example would be the Kac module
  $\cK_\mu$.

  Another important module, even though not
  indecomposable, is the module
  $\cB_\mu=\Ind_{\g_0}^{\g}(V_\mu)$ which is induced from
  a finite dimensional simple $\g_0$-module $V_\mu$. $\cB_\mu$ is
  projective since $V_\mu$ is projective and hence it possesses a
  decomposition
\begin{align}
  \label{eq:BRep}
  \cB_\mu
  \ =\ \bigoplus_{\nu\in\Rep(\g)}m_{\mu\nu}\,\cP_\nu
\end{align}
  into projective covers. Due to the relation
  $\dim\Hom_\g\bigl(\cP_\mu,\cL_\nu\bigr)=\delta_{\mu\nu}$, the
  multiplicities can be obtained as the dimension
\begin{align}
  \label{eq:Multiplicities}
  m_{\mu\nu}
  \ =\ \dim\Hom_\g\bigl(\cB_\mu,\cL_\nu\bigr)
\end{align}
  of a suitable space of $\g$-homomorphisms.

  Finally we introduce an equivalence relation on the set $\Rep(\g)$.
  Two weights $\mu$ and $\nu$ are said to be in the
  same block if there exists a non-split extension
\begin{align}
  0\rightarrow\cL_\mu\rightarrow\cA\rightarrow\cL_\nu\rightarrow0\ \ .
\end{align}
  In other words, if $\cL_\mu$ and $\cL_\nu$ can be obtained as a
  submodule and a quotient of $\cA$, respectively, but nevertheless
  $\cA$ is different from $\cL_\mu\oplus\cL_\nu$. The division of
  weights into blocks defines an equivalence relation. We will use the
  symbol $[\sigma]$ to denote the block $\sigma$ belongs to. A weight
  is typical if and only if the corresponding block contains precisely
  one element. The symbol $\text{AtypBlocks}(\g)$ will be reserved for
  the set of blocks obtained from {\em atypical} modules $\sigma$.

\subsection{Harmonic analysis on supergroups}

  Roughly speaking, a Lie supergroup $\text{G}$ is a fermionic extension of
  an ordinary Lie group $\text{G}_0$ by fermionic coordinates which transform
  in a suitable representation of $\text{G}_0$. All the properties of a
  supergroup are inherited from these data. For this reason, we will
  review the well-known harmonic analysis on ordinary compact Lie
  groups first. The goal of harmonic analysis is to
  learn about the structure of a manifold from studying the action of
  differential operators -- Lie derivatives and Laplace operators --
  on its algebra of functions.

  Let us consider a compact, simple, simply-connected Lie group
  $\text{G}_0$. According to the Peter-Weyl theorem, the algebra
  $\cF(\text{G}_0)$ of square integrable functions on $\text{G}_0$
  (with respect to the Haar measure) decomposes as
\begin{align}
  \cF(\text{G}_0)
  \ \cong\ \bigoplus_{\mu\in\Rep(\text{G}_0)}V_\mu\otimes V_\mu^\ast
\end{align}
  under the left-right regular action $l\times r\cdot f:g\mapsto
  f(l^{-1}gr)$ of $\text{G}_0\times \text{G}_0$, where the sum is over
  all finite dimensional irreducible representations of
  $\text{G}_0$. Each individual term in the decomposition can be
  thought of as being associated with representation matrices
  $\rho_\mu(g)\in\End(V_\mu)$ with $g\in\text{G}_0$. The statement of
  the Peter-Weyl theorem is that these matrix elements can be used to
  approximate any function on $\text{G}_0$ with arbitrary precision.

  The extension to supergroups is straightforward. Compared to an
  ordinary group, a supergroup comes with additional Grassmann algebra
  valued coordinates which generate the exterior algebra
  $\bigwedge(\g_1^\ast)$. This space admits an obvious action of
  $\g_0$ by the Lie bracket or, equivalently, by conjugation with
  elements from $G_0$. The algebra of functions on the supergroup
  $\text{G}$ is the induced module (with respect to the right action
  of $\text{G}_0$)
\begin{align}
  \cF(\text{G})
  \ =\ \Ind_{\g_0}^{\g}\cF(\text{G}_0)
  \ =\ \cF(\text{G}_0)\otimes\bigwedge(\g_1^\ast)\ \ .
\end{align}
  This definition has a natural interpretation arising from formally
  expanding functions on $\text{G}$ in a Taylor series in the odd
  coordinates. Apart from the right action of $\text{G}$, $\cF(\text{G})$ also
  admits a left action, just as in the bosonic case. Our goal is to
  understand the decomposition of this algebra as a
  $\g\oplus\g$-module (with respect to the left and right regular
  action). The result will provide a super-analogue of
  the Peter-Weyl theorem.
  
  Since all finite dimensional representations of a reductive Lie
  algebra $\g_0$ are projective the same will be true for the induced
  module $\cF(\text{G})$. Hence, as a right $\g$-module, $\cF(\text{G})$ has the
  decomposition
\begin{align}
  \label{eq:FGright}
  \cF(\text{G})
  \ =\ \bigoplus_{\mu\in\Rep(G)} L_\mu\otimes\cP^\ast_\mu\ \ ,
\end{align}
  where the sum is over all projective covers of $\g$ and the $L_\mu$
  are some multiplicity spaces. As a left $\g$-module, $\cF(\text{G})$ has
  precisely the same decomposition. Indeed, the algebra of functions
  has to be isomorphic with respect to the left and the right regular
  action due to the existence of the isomorphism
  $\Omega:\cF(\text{G})\to\cF(\text{G})$ which acts as  $\Omega(f):g\mapsto
  f(g^{-1})$ and which intertwines the left and right
  regular actions.

  In the typical sector, $\cP_\mu$ agrees with $\cL_\mu$. Given the
  symmetry between the left and the right action it is then obvious
  that $L_\mu\cong\cL_\mu$ as vector spaces. We will now show that
  this is indeed always the case, not only in the typical sector but
  also in the atypical sector. First of all we notice that, by
  definition and eq.~\eqref{eq:BRep}, the algebra of functions on $\text{G}$
  has the form
\begin{align}
  \cF(\text{G})
  \ =\ \bigoplus_{\mu\in\Rep(\g_0)}V_\mu\otimes\cB_\mu^\ast
  \ =\ \bigoplus_{\mu\in\Rep(\g_0)}m_{\mu\nu}\,V_\mu\otimes\cP_\mu^\ast
\end{align}
  as a $\g_0\oplus\g$-module. We then employ Frobenius reciprocity to
  rewrite eq.~\eqref{eq:Multiplicities} as
\begin{align}
  m_{\mu\nu}
  \ =\ \dim\Hom_{\g_0}\bigl(V_\mu,\cL_\nu\bigr)\ \ ,
\end{align}
  which proves our assertion, given that all $\g_0$-modules are fully
  reducible.

  The result just obtained suggests
  that we have a factorization $\cL_\mu\otimes\cL^\ast_\mu$ of the
  individual contributions in the typical sector, just as in the
  case of $\cF(\text{G}_0)$. In the atypical sector, however, such a
  factorization is not possible since the projective covers $\cP_\mu$
  are strictly larger than the simple modules $\cL_\mu$. For this
  reason, the left and right modules in the atypical sector are
  entangled in a complicated way, arranging themselves in infinite
  dimensional non-chiral indecomposable $\g\oplus\g$-modules
  $\cI_{[\sigma]}$ for each individual {\em block} $[\sigma]$. We
  finally find\footnote{Versions of this result appear to be known among
    mathematicians specialized on Lie superalgebras, even though no
    specific reference seems to exist. In the physics literature the
    result was first noted in \cite{Quella:2007hr}.}
\begin{align}
  \cF(\text{G})
  \ =\ \bigoplus_{\mu\in\Typ(\g)}\cL_\mu\otimes\cL^\ast_\mu
       \oplus\bigoplus_{[\sigma]\in\text{AtypBlocks}(\g)}\cI_{[\sigma]}
\end{align}
  for the decomposition of $\cF(\text{G})$ as a $\g\oplus\g$-module. For Lie
  supergroups of type I this has been worked out in great detail in
  \cite{Quella:2007hr}.

  It should be emphasized that the Laplace operator is not
  diagonalizable on the atypical projective covers $\cP_\mu$ and on
  the non-chiral modules $\cI_{[\sigma]}$. For $\sigma$-models on
  supergroups this implies that they are generally logarithmic
  conformal field theories. Exceptions may occur for small volumes
  where the spectrum can be truncated in such a way that the modules
  $\cI_{[\sigma]}$ no longer contribute.

\begin{figure}
\begin{center}
\begin{pgfpicture}{0cm}{0.2cm}{13cm}{2cm}
  \pgfputat{\pgfxy(1.4,2.2)}{\pgfbox[center,center]{a) Typical sector}}
  \pgfputat{\pgfxy(7.8,2.2)}{\pgfbox[center,center]{b) Atypical sector}}
  \pgfputat{\pgfxy(0.9,0)}{\pgfbox[center,center]{\tiny Left action}}
  \pgfputat{\pgfxy(2.9,0)}{\pgfbox[center,center]{\tiny Right action}}
  \pgfputat{\pgfxy(4.8,0)}{\pgfbox[center,center]{\tiny Both actions}}
  \pgfputat{\pgfxy(7.2,0)}{\pgfbox[center,center]{\tiny Left action}}
  \pgfputat{\pgfxy(9.2,0)}{\pgfbox[center,center]{\tiny Right action}}
  \pgfputat{\pgfxy(11.1,0)}{\pgfbox[center,center]{\tiny Both actions}}
\begin{pgfmagnify}{0.9}{0.9}
  \pgfputat{\pgfxy(3,1)}{\pgfbox[center,center]{\pgfimage[width=6cm]{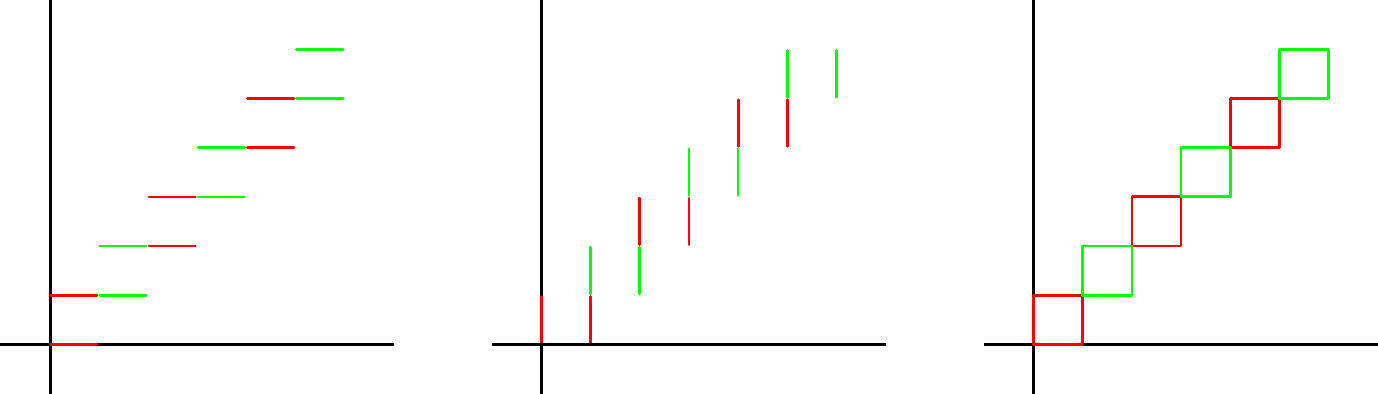}}}
  \pgfputat{\pgfxy(10,1)}{\pgfbox[center,center]{\pgfimage[width=6cm]{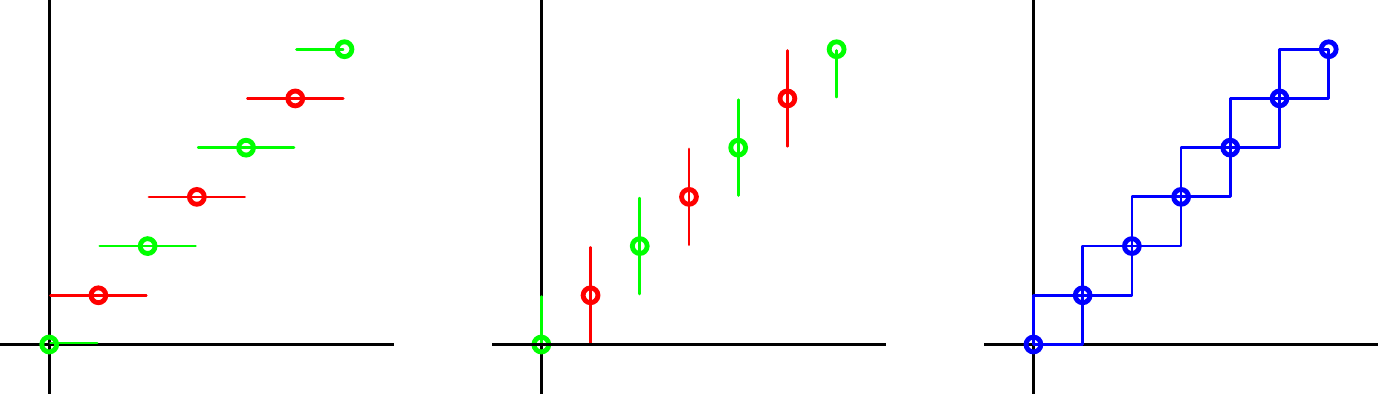}}}
\end{pgfmagnify}
\end{pgfpicture}
\end{center}
  \caption{\label{fig:GL}(Color online) Sketch of the harmonic
    analysis on supergroups using the example of $\text{GL}(1|1)$. It
    is shown how the space of functions organizes itself with respect
    to the left and the right action of $\g$ (the axes correspond to
    eigenvalues of the respective Cartan generators) and with respect
    to the simultaneous action of both. In the typical sector a) we
    observe a factorization of representation (green and red) while in
    the atypical sector b) the functions organize themselves in
    infinite dimensional non-factorizing representations (blue) due to
    the extension of simple modules into projective covers.}
\end{figure}

\subsection{Example:  \texorpdfstring{$\text{GL}(1|1)$}{GL(1|1)}}

  Since our previous result has been very abstract, let us explain it
  in more detail using the example of $\text{GL}(1|1)$ (first considered in
  \cite{Schomerus:2005bf}). The corresponding Lie superalgebra
  $\text{gl}(1|1)$ has two Cartan elements $E$ and $N$. Consequently, states
  in a module are
  labeled by two numbers $(e,n)$, the eigenvalues of $E$ and $N$.
  Since $E$ is central, the value $e$ merely plays the role
  of a spectator. For $e=0$ the state sits in an atypical
  representation, otherwise in a typical one. Typical representations
  $\cL(e,n)$ (with $e\neq0$) are two-dimensional while
  atypical simple modules $\cL(n)$ are one-dimensional. On
  the other hand, the projective cover $\cP(n)$  of the atypical simple
  module $\cL(n)$ is four-dimensional, having a composition series
  $\cP(n):\cL(n)\to\cL(n+1)\oplus\cL(n-1)\to\cL(n)$.

  The most important features of the harmonic analysis on $\text{GL}(1|1)$
  are sketched in figure \ref{fig:GL}.
  In the typical sector, the states organize themselves into
  two-dimensional representations, both with respect to the left and
  with respect to the right action (red and green lines). Under the combined
  action they combine into four-dimensional representations (red and green
  boxes) which correspond to the tensor product of two
  representations. In the atypical sector, however, the picture is
  very different. Here the states organize themselves in
  four-dimensional projective covers if only one of the two actions is
  considered (red and green lines). One should imagine
  a diamond which is perpendicular to the plane, with two of the
  vertices being located in the plane. In this case the states cannot
  be organized in a tensor product under the combined action for
  obvious reasons. Instead they have to combine into infinite
  dimensional indecomposable multiplets (blue), one for each value of
  $n\text{ mod }1$.

\subsection{Harmonic analysis on supercosets}
  
  The harmonic analysis on a supercoset $\text{G/H}$ where the
  supergroup elements are identified according to the rule $g\sim gh$
  with $h\in\text{H}$ can be immediately deduced from that of the
  supergroup case. Indeed, the algebra of functions on $\text{G/H}$
  can be thought of as the space of $\text{H}$-invariant functions on
  $\text{G}$,
\begin{align}
  \label{eq:InvSpace}
  \cF(\text{G/H})
  \ =\ \Inv_H\cF(\text{G})\ \ ,
\end{align}
  where an element $h$ of the supergroup $\text{H}$ acts on
  $f\in\cF(\text{G})$ according to
\begin{align}
  h\cdot f(g)
  \ =\ f(gh)\ \ .
\end{align}
  It is obvious that the space $\text{G/H}$ and hence also the algebra
  of functions $\cF(\text{G/H})$ still admits an action of
  $\text{G}$. The isometry supergroup of $\text{G/H}$ might be bigger
  than $\text{G}$ but for simplicity we will only consider the
  symmetry $\text{G}$.

  Writing the invariant subspace \eqref{eq:InvSpace} explicitly as a
  direct sum over indecomposable $\text{G}$-modules turns out to be rather
  involved in the general case. The main reason is that the modules
  over Lie superalgebras are not fully decomposable. On the one hand
  such modules already appear in $\cF(\text{G})$, as
  $\text{G}$-modules with respect
  to the right regular action, see eq.~\eqref{eq:FGright}. On the
  other hand, they may also arise when decomposing simple $\text{G}$-modules
  after restricting the action to the supergroup $\text{H}$. Finally, the
  invariants that need to be extracted when restricting from $\cF(\text{G})$
  to $\cF(\text{G/H})$ can be either true $\text{H}$-invariants of $\cF(\text{G})$ (i.e.\
  simple $\text{H}$-modules) or they can sit in a larger indecomposable
  $\text{H}$-module. A general solution to this intricate problem is
  currently beyond reach. Nevertheless we can state one general
  lesson: In case $\text{H}$ is purely bosonic the algebra of functions
  $\cF(\text{G/H})$ will necessarily involve projective covers of simple
  $\text{G}$-modules. In particular, this observation is relevant for the
  supercosets describing $\text{AdS}$ backgrounds in string theory, see table
  \ref{tab:Strings}.

  The problem sketched in the previous paragraph can be circumvented
  when working on the level of characters since these are not
  sensitive to the indecomposable structure of modules. For our
  purposes this will be sufficient.

\subsection{Example: The supersphere  \texorpdfstring{$\text{S}^{3|2}$}{S(3|2)}}

  Instead of treating the general case, we discuss one example in
  some detail, namely the supersphere $\text{S}^{3|2}$. This supersphere can
  be thought of as being embedded into the flat superspace
  $\Real^{4|2}$, i.e.\ we have four bosonic coordinates $x^i$ and two
  fermionic coordinates $\eta_1,\eta_2$ subject to the constraint
  $\vec{x}^2+\eta_1\eta_2=R^2$. For our purposes we can identify the
  algebra of functions $\cF(\text{S}^{3|2})$ with the polynomial algebra in
  the six coordinates $X^a\in\Real^{4|2}$ modulo the ideal
  $\vec{X}^2=R^2$.

  These coordinates transform in the vector representation of
  $\text{SO}(4)$ and the defining representation of $\text{SP}(2)$,
  respectively. Moreover, these transformations leave the constraint
  invariant. In addition, one can consider transformations which mix
  bosons and fermions. The resulting supergroup of isometries is
  $\text{OSP}(4|2)$. Since the stabilizer of an arbitrary point on
  $\text{S}^{3|2}$ is $\text{OSP}(3|2)$, this confirms that the
  supersphere possesses a representation as a supercoset
  $\text{OSP}(4|2)/\text{OSP}(3|2)$ (cf.\ table \ref{tab:CondMat}).

  In order to determine the character for the $\text{OSP}(4|2)$-module
  $\cF(\text{S}^{3|2})$ we proceed as follows. We first identify the Cartan
  subalgebra of $\text{osp}(4|2)$ with the Cartan subalgebra of
  $\text{su}(2)\times\text{su}(2)\times\text{su}(2)$, where we use the
  identification $\text{so}(4)\cong\text{su}(2)\times\text{su}(2)$ and
  $\text{sp}(2)\cong\text{su}(2)$. We then choose linear
  combinations of the six coordinates such that we can assign the
  weights $(\epsilon,\eta,0)$ and $(0,0,\epsilon)$ to
  them, with $\epsilon,\eta=\pm1$. In
  addition we introduce a quantum number for the polynomial
  grading. Each of the six coordinates $X^a$ then contributes a term
  $z_1^{m_1}z_2^{m_2}z_3^{m_3}t$ to the character, where
  $m_i\in\{0,\pm1\}$ have to be chosen according to the respective
  weights and $t$ keeps track of the polynomial grade. For a
  product of coordinates, these individual contributions have to be
  multiplied with each other since the quantum numbers and the
  polynomial degree add up.

  After these preparations it is very simple to write down the
  character of all polynomial functions in the coordinates
  $X^a$. Dividing out the ideal $\vec{X}^2=R^2$ is taken into account
  by multiplying the previous character with $1-t^2$ since the
  constraint relates singlets at polynomial degree $n$ to singlets
  at polynomial degree $n-2$. At the end we take the limit $t\to1$
  since the grade is not a good quantum number once we impose the
  constraint. The total character is thus given by
\begin{align}
  \label{eq:ZSS1}
  Z_{\cF(\text{S}^{3|2})}(z_1,z_2,z_3)
  \ =\ \lim_{t\to1}\frac{(1-t^2)(1+tz_3)(1+t/z_3)}{(1-tz_1z_2)(1-tz_1/z_2)(1-tz_2/z_1)(1-t/z_1z_2)}\ \ .
\end{align}
  This expression can be expanded and represented as a linear
  combination of characters of $\text{osp}(4|2)$. Without going into the
  details we just write down the result
\begin{align}
  Z_{\cF(\text{S}^{3|2})}(z_1,z_2,z_3)
  \ =\ \chi_{[0,0,0]}(z_1,z_2,z_3)
       +\sum_{k=0}^\infty\chi_{[1/2,k/2,k/2]}(z_1,z_2,z_3)\
  \ ,
\end{align}
  where the labels $[j_1,j_2,j_3]$ refer to simple modules (see
  \cite{Mitev:2008yt}). Since
  $[0,0,0]$ denotes the trivial representation and all the other
  labels correspond to typical representations, this character
  decomposition at the same time yields the result for the harmonic
  analysis on $\text{S}^{3|2}$. The considerations of this section will be
  used in section \ref{sc:SphereSpectrum} below when we discuss the
  spectrum of freely moving open strings on $\text{S}^{3|2}$.

\section{\label{sc:Deformations}Supergroup WZW models and their deformations}

  A special class of conformal superspace $\sigma$-models are WZW
  models. They are distinguished by the existence of an infinite
  dimensional affine Kac-Moody superalgebra symmetry. We review
  supergroup WZW models and discuss two deformations which are special
  for supergroups with vanishing Killing form.

\subsection{\label{sc:WZW}WZW models}

  Let us fix a supergroup $\text{G}$ and a non-degenerate invariant bilinear
  form $\langle\cdot,\cdot\rangle$. We assume the supergroup to be
  simple and simply-connected and the invariant form to be normalized in the
  standard way (see below). The supergroup WZW model is a
  two-dimensional $\sigma$-model describing the propagation of strings
  on $\text{G}$. The action functional is given by
\begin{align}
  \label{eq:WZW}
  \cS^{\text{WZW}}[g]
  \ =\ -\frac{k}{2\pi}\int_\Sigma\!d^2z\,
       \langle g^{-1}\partial g,g^{-1}\bartial g\rangle
       -\frac{ik}{24\pi}\,\int_B
       \langle g^{-1}dg,[g^{-1}dg,g^{-1}dg]\rangle\ \ ,
\end{align}
  where $\Sigma$ is a closed Riemann surface and $B$ is a
  three-dimensional extension of this surface such that $\partial
  B=\Sigma$. The form $\langle\cdot,\cdot\rangle$ is supposed to be
  normalized such that the topological Wess-Zumino term is
  well-defined up to multiples of $2\pi i$ as long as $k$ is an
  integer. The level $k$ is thus the only parameter of the
  model.\footnote{For simple Lie supergroups all invariant forms are
    unique up to rescaling.}

  By construction, every WZW model has a global symmetry $\text{G}\times\text{G}$
  corresponding to multiplying the field $g(z,\bz)$ by arbitrary group
  elements from the left and from the right. In fact, this symmetry is
  elevated to an affine Kac-Moody superalgebra symmetry
  $\widehat{\text{G}}_k$
\begin{align}
  \label{eq:WZWOPE}
  J^a(z)\,J^b(w)
  &\ =\ \frac{k\,\kappa^{ab}}{(z-w)^2}
        +\frac{i{f^{ab}}_c\,J^c(w)}{z-w}
        +\text{non-singular}
\end{align}
  and a corresponding anti-holomorphic symmetry if one allows these
  elements to depend holomorphically and antiholomorphically on $z$,
  respectively. In the last formula, the currents are defined by
  $J=-k\partial gg^{-1}$ and $\bJ=kg^{-1}\bartial g$. The equations
  of motion guarantee that they are holomorphic and antiholomorphic,
  respectively.

\subsection{\label{sc:WZWDef1} \texorpdfstring{$\text{G}\times\text{G}$}{GxG} preserving deformations}

\begin{figure}
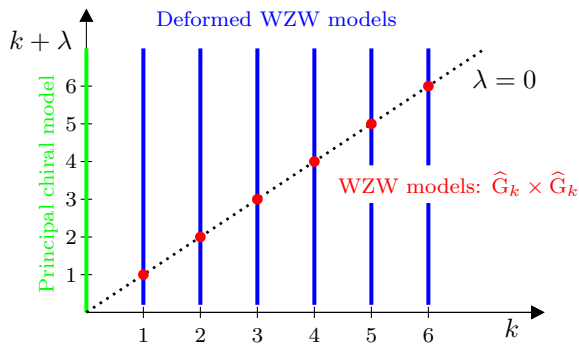

\begin{center}
\begin{pgfpicture}{0cm}{0cm}{6cm}{4cm}
  \pgfsetendarrow{\pgfarrowtriangle{4pt}}
  \pgfxyline(0,0)(6,0)
  \pgfxyline(0,0)(6,0)
  \pgfxyline(0,0)(0,4)
  \pgfclearendarrow
  \pgfxyline(.75,-.075)(.75,.075)
  \pgfxyline(1.5,-.075)(1.5,.075)
  \pgfxyline(2.25,-.075)(2.25,.075)
  \pgfxyline(3,-.075)(3,.075)
  \pgfxyline(3.75,-.075)(3.75,.075)
  \pgfxyline(4.5,-.075)(4.5,.075)
  \pgfxyline(-.075,.5)(.075,.5)
  \pgfxyline(-.075,1)(.075,1)
  \pgfxyline(-.075,1.5)(.075,1.5)
  \pgfxyline(-.075,2)(.075,2)
  \pgfxyline(-.075,2.5)(.075,2.5)
  \pgfxyline(-.075,3)(.075,3)
  \pgfputat{\pgfxy(5.6,-.2)}{\pgfbox[center,center]{$k$}}
  \pgfputat{\pgfxy(-.6,3.6)}{\pgfbox[center,center]{$k+\lambda$}}
  \pgfputat{\pgfxy(5.5,3.1)}{\pgfbox[center,center]{$\lambda=0$}}
  \pgfputat{\pgfxy(-.2,.5)}{\pgfbox[center,center]{\footnotesize1}}
  \pgfputat{\pgfxy(-.2,1)}{\pgfbox[center,center]{\footnotesize2}}
  \pgfputat{\pgfxy(-.2,1.5)}{\pgfbox[center,center]{\footnotesize3}}
  \pgfputat{\pgfxy(-.2,2)}{\pgfbox[center,center]{\footnotesize4}}
  \pgfputat{\pgfxy(-.2,2.5)}{\pgfbox[center,center]{\footnotesize5}}
  \pgfputat{\pgfxy(-.2,3)}{\pgfbox[center,center]{\footnotesize6}}
  \pgfputat{\pgfxy(.75,-.3)}{\pgfbox[center,center]{\footnotesize1}}
  \pgfputat{\pgfxy(1.5,-.3)}{\pgfbox[center,center]{\footnotesize2}}
  \pgfputat{\pgfxy(2.25,-.3)}{\pgfbox[center,center]{\footnotesize3}}
  \pgfputat{\pgfxy(3,-.3)}{\pgfbox[center,center]{\footnotesize4}}
  \pgfputat{\pgfxy(3.75,-.3)}{\pgfbox[center,center]{\footnotesize5}}
  \pgfputat{\pgfxy(4.5,-.3)}{\pgfbox[center,center]{\footnotesize6}}
  \pgfsetlinewidth{1.5pt}
  {\color{blue}\pgfxyline(.75,.1)(.75,3.5)}
  {\color{blue}\pgfxyline(1.5,.1)(1.5,3.5)}
  {\color{blue}\pgfxyline(2.25,.1)(2.25,3.5)}
  {\color{blue}\pgfxyline(3,.1)(3,3.5)}
  {\color{blue}\pgfxyline(3.75,.1)(3.75,1.45)}
  {\color{blue}\pgfxyline(3.75,1.95)(3.75,3.5)}
  {\color{blue}\pgfxyline(4.5,.1)(4.5,1.45)}
  {\color{blue}\pgfxyline(4.5,1.95)(4.5,3.5)}
  {\color{green}\pgfxyline(0,0)(0,3.5)}
  \pgfsetlinewidth{1pt}
  \pgfsetdash{{1pt}{2pt}}{0pt}  
  \pgfxyline(0,0)(5.25,3.5)
  {\red\pgfcircle[fill]{\pgfxy(.75,.5)}{2pt}}
  {\red\pgfcircle[fill]{\pgfxy(1.5,1)}{2pt}}
  {\red\pgfcircle[fill]{\pgfxy(2.25,1.5)}{2pt}}
  {\red\pgfcircle[fill]{\pgfxy(3,2)}{2pt}}
  {\red\pgfcircle[fill]{\pgfxy(3.75,2.5)}{2pt}}
  {\red\pgfcircle[fill]{\pgfxy(4.5,3)}{2pt}}
  \begin{pgfrotateby}{\pgfpoint{1pt}{0pt}}
    {\color{green}\pgfputat{\pgfxy(1.7,.5)}{\pgfbox[center,center]{\footnotesize
          Principal chiral model}}}
  \end{pgfrotateby}
  {\red\pgfputat{\pgfxy(4.9,1.7)}{\pgfbox[center,center]{\footnotesize
        WZW models: $\widehat{\text{G}}_k\times\widehat{\text{G}}_k$}}}
  {\color{blue}\pgfputat{\pgfxy(2.5,3.9)}{\pgfbox[center,center]{\footnotesize Deformed WZW models}}}
\end{pgfpicture}
\end{center}
  \caption{\label{sc:GGModuliSpace}(Color online) The moduli space of
    $\text{G}\times\text{G}$ preserving deformations of WZW models on
    supergroups with vanishing Killing form. The WZW models and the
    principal chiral models are associated with the special loci
    $\lambda=0$ and $k=0$, respectively.}
\end{figure}

  WZW models possess an obvious deformation which amounts to allowing
  for a different normalization of the two contributions to the action
  \eqref{eq:WZW}. Since the coefficient of the topological term is
  quantized (at least for supergroups having a compact simple part),
  the only freedom is in fact to change the coefficient of the kinetic
  term. In most cases, especially for all simple bosonic groups, such
  a deformation would spoil conformal invariance. The only exception
  occurs if $\text{G}$ is a supergroup with vanishing Killing form as
  we will now review.

  The implementation of the $\text{G}\times\text{G}$-preserving
  deformation in the operator language is somewhat cumbersome. We have
  to find a field $\Phi(z,\bz)$ that is invariant under
  $\text{G}\times\text{G}$ and possesses conformal dimension
  $(h,\bh)=(1,1)$. The deformed model would then formally be described
  by the deformed action
  $\cS_{\text{def}}=\cS^{\text{WZW}}+\lambda/2\pi\int\!d^2z\,\Phi(z,\bz)$. The
  two currents $J$ and $\bJ$ transform non-trivially under either the
  left or the right action of $\text{G}$ on itself. For this reason,
  an invariant can only be built by conjugating one of the two
  currents. In algebraic terms we have to consider the normal ordered
  operator
\begin{align}
  \Phi_1(z,\bz)
  \ =\ :\bigl\langle J(z),\Ad_g\bigl(\bJ(\bz)\bigr)\bigr\rangle:
  \ =\ :J^a\phi_{ab}\bJ^b:(z,\bz)
\end{align}
  involving the non-chiral affine primary field $\phi_{ab}(z,\bz)$
  which transforms in the representation $\ad\times\ad$ with respect
  to the symmetry $\text{G}\times\text{G}$. At lowest order in perturbation theory,
  the field $\Phi_1(z,\bz)$ is marginal if and only if the field
  $\phi_{ab}$ has conformal dimensions $(h,\bh)=(0,0)$. Since the
  conformal dimensions of the affine primary field $\phi_{\mu\nu}$ are
  proportional to the quadratic Casimir in the adjoint representation,
  this condition is equivalent to the vanishing of the Killing form.

  A sketch of the moduli spaces for conformal supergroup
  $\sigma$-models and the symmetries at each individual point can be
  found in figure \ref{sc:GGModuliSpace}. In case $\text{G}$ is bosonic (but
  not abelian) or its Killing form is not vanishing, conformal
  invariance only holds for WZW models. The moduli space of
  $\text{G}\times\text{G}$ symmetric models is discrete in that case.
  WZW models possess a special kind of D-branes preserving the
  diagonal $\text{G}$ symmetry. The simplest of these is a point-like D-brane
  which is localized at the identity element of $\text{G}$. The spectrum of
  anomalous dimensions on such a D-brane has been determined in
  \cite{Quella:2007sg},
\begin{align}
  \label{eq:GGSpectrum}
  \delta h_\Lambda
  \ =\ -\frac{k\lambda}{1+k\lambda}\,\frac{C_\Lambda}{k}\ \ .
\end{align}
  Here the label refers to a field which transforms in the
  representation $\Lambda$ with respect to the global $\text{G}$-symmetry and
  $C_\Lambda$ is the corresponding eigenvalue of the quadratic Casimir.

  The analysis above is directly relevant for the study of strings on
  $\text{AdS}_3\times\text{S}^3$ \cite{Berkovits:1999im}. The coefficients
  $\lambda$ and $k$ are in one-to-one correspondence to the number of
  RR and NS fluxes in the background. Our formula
  \eqref{eq:GGSpectrum} provides the first exact string spectrum on
  $\text{AdS}_3\times\text{S}^3$ with non-trivial RR flux.

\subsection{\label{sc:WZWDef2} \texorpdfstring{$\text{G}$}{G} preserving deformations}

  Supergroups with vanishing Killing form admit a second type of
  deformation. This deformation is easier to describe since it avoids
  the use of the cumbersome field $\phi_{\mu\nu}$. On the other hand
  it is more difficult to interpret geometrically since the group
  structure is broken. In this case, the deformation is implemented by
  the perturbing field
\begin{align}
  \label{eq:Deformation}
  \Phi_2(z,\bz)
  \ =\ \,:\!\bigl\langle J(z),\bJ(\bz)\bigr\rangle\!:
  \ =\ \,:\!J_a\bJ^a(z,\bz)\!:\ \ .
\end{align}
  This field transforms non-trivially under the full
  $\text{G}\times\text{G}$-symmetry but trivially under the diagonal subgroup $\text{G}$. As a
  consequence, the original global $\text{G}\times\text{G}$-symmetry is broken to
  the diagonal $\text{G}$-symmetry as soon as the deformation is switched
  on. In the present case it is less obvious than for $\Phi_1(z,\bz)$
  but again we need to impose the vanishing of the Killing form since
  otherwise the field $\Phi_2(z,\bz)$ would break conformal invariance
  at higher orders in perturbation theory.

  The deformed model admits D-branes which preserve the full global
  $\text{G}$-symmetry. They are obtained from symmetry preserving D-branes in
  the WZW model that preserve one copy of the current algebra
  $\widehat{\text{G}}_k$. For open strings ending on such D-branes, the anomalous
  dimensions can be determined explicitly using conformal perturbation
  theory \cite{Mitev:2008yt}. We postpone the explanation to section
  \ref{sc:QuasiAbelian} and just state the result
\begin{align}
  \label{eq:AnomalousDimension2}
  \delta h_\Lambda
  \ =\ -\frac{k\lambda}{1+k\lambda}\,\frac{C_\Lambda}{k}\ \ .
\end{align}
  The agreement of this expression with \eqref{eq:GGSpectrum}
  is no coincidence since both have their origin in the same kind of
  combinatorics. Again, $C_\Lambda$ refers to the eigenvalue of
  the quadratic Casimir acting on the representation $\Lambda$.

\section{A duality between Gross-Neveu models and supersphere  \texorpdfstring{$\sigma$}{sigma}-models}

  In this section, we present arguments which support a conjectured
  duality between $\text{OSP}(2S+2|2S)$ Gross-Neveu models and
  $\sigma$-models on the superspheres $\text{S}^{2S+1|2S}$. We first
  point out the formulation of the Gross-Neveu model as a deformed WZW
  model and how boundary spectra can be calculated along the lines of
  section \ref{sc:Deformations}. For simplicity we restrict ourselves
  to the case of $S=1$.

\subsection{The  \texorpdfstring{$\text{OSP}(2S+2|2S)$}{OSP(2S+2|2S)} Gross-Neveu model as a deformed WZW model}

  The results of the previous section may be used to present arguments
  in favor of a duality between non-linear $\sigma$-models on
  superspheres $\text{S}^{2S+1|2S}$ and $\text{OSP}(2S+2|2S)$ Gross-Neveu models
  \cite{Mitev:2008yt}. In
  the case $S=0$, this duality reduces to the well-known correspondence
  between the massless Thirring model (also known as Luttinger liquid
  in the condensed matter community) and the free compactified
  boson. All cases $S\geq1$ can be thought of as non-abelian
  generalizations of this equivalence.

  The $\text{OSP}(4|2)$ Gross-Neveu model is a non-geometric theory defined
  by the following Lagrangian:
\begin{align}
  \cS^{\text{GN}}[\Psi]
  &\ =\ \frac{1}{2\pi}\int
  d^2z\Bigl[\langle\Psi,\bartial\Psi\rangle+\langle\bar{\Psi},\partial\bar{\Psi}\rangle+g^2\langle\Psi,\bar{\Psi}\rangle^2\Bigr]\ \ .
\end{align}
  Here, $\Psi=(\psi_1,\ldots,\psi_4,\beta,\gamma)$ is a fundamental
  $\text{OSP}(4|2)$ multiplet with four fermions and two bosons, all having
  conformal dimension $h=1/2$.
  The theory has a single coupling constant $g$ which determines the
  strength of the quadratic potential. Instead of working with
  the $\text{OSP}(4|2)$ Gross-Neveu model it is convenient to use the
  components of the free fields $\Psi^i$ to construct the
  currents of an $\widehat{\text{OSP}}(4|2)$ affine Lie
  superalgebra at level $k=1$. Using these currents,
  the previous Lagrangian can be written as a deformed WZW model
\begin{align}
  \cS^{\text{GN}}
  &\ =\ \cS^{\text{WZW}}+g^2\cS_{\text{def}}&
  &\text{with}&
  \cS_{\text{def}}
  &\ =\ \frac{1}{2\pi}\int\!d^2z\,\bigl\langle J,\omega(\bJ)\bigr\rangle\ \ ,
\end{align}
  $\omega$ being induced from the exchange automorphism of
  $\widehat{\text{SU}}(2)_1\times\widehat{\text{SU}}(2)_1$. This kind
  of deformation is covered by our discussion in section
  \ref{sc:WZWDef2}.

  The solution of the $\text{OSP}(4|2)$ WZW model is relatively
  straightforward, since it can simply be formulated as an orbifold
\begin{align}
  \widehat{\text{OSP}}(4|2)_1
  \ =\ \Bigl(\widehat{\text{SU}}(2)_{-\frac{1}{2}}\times\widehat{\text{SU}}(2)_1\times
  \widehat{\text{SU}}(2)_1\Bigr)\Bigr/\Integer_2
\end{align}
  of purely bosonic WZW models.\footnote{The orbifold corresponds to
    the action of the simple current $(1/2,1/2,1/2)$. It implements a
    target space GSO projection.} The two copies of $\text{SU}(2)_1$
  arise from the two pairs of fermions, while the
  $\widehat{\text{SU}}(2)_{-\frac{1}{2}}$ arises from the bosonic
  $\beta\gamma$ system. Both theories are well-understood, even though
  the $\beta\gamma$ system exhibits some rather unexpected features
  \cite{Lesage:2002ch}. It is worth noting that despite the
  subtleties, the WZW model on $\text{OSP}(4|2)$ at level $k=1$ is not
  a logarithmic CFT, in contrast to WZW models at higher levels
  \cite{Schomerus:2005bf,Quella:2007hr}. This is due to the
  realization in terms of free fields.\footnote{However, in close
    analogy to \cite{Lesage:2003kn}, it is possible to construct a
    logarithmic lift of this theory by including fermionic
    zero-modes.}

  The D-brane we wish to study corresponds to trivial gluing
  conditions in the $\widehat{\text{SU}}(2)_{-\frac{1}{2}}$ part and
  permutation gluing conditions in the
  $\widehat{\text{SU}}(2)_1\times\widehat{\text{SU}}(2)_1$ part. Its
  spectrum can easily be determined to be \cite{Mitev:2008yt}
\begin{align}
  \label{eq:WZWSpectrum}
  Z_{\text{GN}}(q,z|0)
  &\ =\ \frac{\eta(q)}{\theta_4(z_1)}\Biggl[\frac{\theta_2(q^2,z_2^2)\theta_2(q^2,z_3^2)}{\eta(q)^2}
        +\frac{\theta_3(q^2,z_2^2)\theta_3(q^2,z_3^2)}{\eta(q)^2}\Biggr]\
        \ .
\end{align}
  It is just the sum of the affine $\widehat{\text{OSP}}(4|2)_1$
  characters based on the trivial and the fundamental representation
  of $\text{OSP}(4|2)$.

\subsection{Deformed boundary spectrum}

\begin{figure}
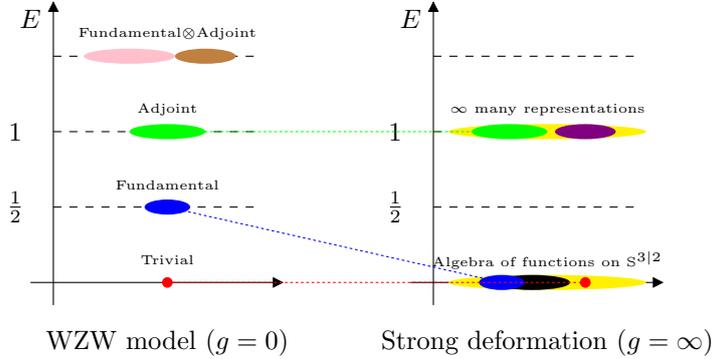

\begin{center}
\begin{pgfpicture}{0cm}{0cm}{10cm}{5cm}
  \pgfsetendarrow{\pgfarrowtriangle{3pt}}
  \pgfline{\pgfxy(.7,.8)}{\pgfxy(4,.8)}
  \pgfline{\pgfxy(5.7,.8)}{\pgfxy(9,.8)}
  \pgfline{\pgfxy(1,0.5)}{\pgfxy(1,4.5)}
  \pgfline{\pgfxy(6,0.5)}{\pgfxy(6,4.5)}
  \pgfputat{\pgfxy(2.5,0)}{\pgfbox[center,center]{WZW model ($g=0$)}}
  \pgfputat{\pgfxy(7.5,0)}{\pgfbox[center,center]{Strong deformation ($g=\infty$)}}
  \pgfputat{\pgfxy(0.7,4.3)}{\pgfbox[center,center]{$E$}}
  \pgfputat{\pgfxy(0.5,1.8)}{\pgfbox[center,center]{$\frac{1}{2}$}}
  \pgfputat{\pgfxy(0.5,2.8)}{\pgfbox[center,center]{$1$}}
  \pgfputat{\pgfxy(5.7,4.3)}{\pgfbox[center,center]{$E$}}
  \pgfputat{\pgfxy(5.5,1.8)}{\pgfbox[center,center]{$\frac{1}{2}$}}
  \pgfputat{\pgfxy(5.5,2.8)}{\pgfbox[center,center]{$1$}}
  \pgfsetdash{{3pt}{3pt}}{0pt}  
  \pgfclearendarrow
  \pgfline{\pgfxy(1,1.8)}{\pgfxy(3.7,1.8)}
  \pgfline{\pgfxy(6,1.8)}{\pgfxy(8.7,1.8)}
  \pgfline{\pgfxy(1,2.8)}{\pgfxy(3.7,2.8)}
  \pgfline{\pgfxy(6,2.8)}{\pgfxy(8.7,2.8)}
  \pgfline{\pgfxy(1,3.8)}{\pgfxy(3.7,3.8)}
  \pgfline{\pgfxy(6,3.8)}{\pgfxy(8.7,3.8)}
  {\red\pgfcircle[fill]{\pgfxy(2.5,.8)}{2pt}}
  {\color{blue}\pgfellipse[fill]{\pgfxy(2.5,1.8)}{\pgfxy(.3,0)}{\pgfxy(0,.1)}}
  {\color{green}\pgfellipse[fill]{\pgfxy(2.5,2.8)}{\pgfxy(.5,0)}{\pgfxy(0,.1)}}
  {\color{brown}\pgfellipse[fill]{\pgfxy(3,3.8)}{\pgfxy(.4,0)}{\pgfxy(0,.1)}}
  {\color{pink}\pgfellipse[fill]{\pgfxy(2,3.8)}{\pgfxy(.6,0)}{\pgfxy(0,.1)}}
  \pgfputat{\pgfxy(2.5,3.1)}{\pgfbox[center,center]{\tiny Adjoint}}
  \pgfputat{\pgfxy(2.5,2.1)}{\pgfbox[center,center]{\tiny Fundamental}}
  \pgfputat{\pgfxy(2.5,1.1)}{\pgfbox[center,center]{\tiny Trivial}}
  \pgfputat{\pgfxy(2.5,4.1)}{\pgfbox[center,center]{\tiny Fundamental$\otimes$Adjoint}}
  {\color{yellow}\pgfellipse[fill]{\pgfxy(7.5,.8)}{\pgfxy(1.3,0)}{\pgfxy(0,.1)}}
  {\color{yellow}\pgfellipse[fill]{\pgfxy(7.5,2.8)}{\pgfxy(1.3,0)}{\pgfxy(0,.1)}}
  {\violet\pgfellipse[fill]{\pgfxy(8,2.8)}{\pgfxy(.4,0)}{\pgfxy(0,.1)}}
  {\color{green}\pgfellipse[fill]{\pgfxy(7.0,2.8)}{\pgfxy(.5,0)}{\pgfxy(0,.1)}}
  {\color{black}\pgfellipse[fill]{\pgfxy(7.3,.8)}{\pgfxy(.5,0)}{\pgfxy(0,.1)}}
  {\color{blue}\pgfellipse[fill]{\pgfxy(6.9,.8)}{\pgfxy(.3,0)}{\pgfxy(0,.1)}}
  {\red\pgfcircle[fill]{\pgfxy(8,.8)}{2pt}}
  \pgfputat{\pgfxy(7.5,1.1)}{\pgfbox[center,center]{\tiny Algebra of
      functions on $\text{S}^{3|2}$}}
  \pgfputat{\pgfxy(7.5,3.1)}{\pgfbox[center,center]{\tiny $\infty$
      many representations}}
  \pgfsetdash{{1pt}{1pt}}{0pt}  
  {\red\pgfline{\pgfxy(2.5,.8)}{\pgfxy(8,.8)}}
  {\color{green}\pgfline{\pgfxy(2.5,2.8)}{\pgfxy(7.0,2.8)}}
  {\color{blue}\pgfline{\pgfxy(2.5,1.8)}{\pgfxy(6.9,.8)}}
\end{pgfpicture}
\end{center}
  \caption{\label{fig:Spectrum}(Color online) A distinguished boundary
    spectrum of the $\text{OSP}(4|2)$ Gross-Neveu model at zero and
    infinite coupling. The interpolation between these two spectra for
    other values of the coupling $g$ is described by formula
    \eqref{eq:Interpolation}.}
\end{figure}

  Once the WZW model has been solved, it is straightforward to
  determine the deformed boundary spectrum using the results of
  section \ref{sc:WZWDef2}. According to formula
  \eqref{eq:AnomalousDimension2} the anomalous dimensions of a
  boundary field only depend on the transformation properties with
  respect to the global $\text{OSP}(4|2)$-symmetry. Taking care of the proper
  normalizations, the partition function then reads\footnote{We use a
    different normalization of the Casimir here.}
\begin{align}
  \label{eq:Interpolation}
  Z_{\text{GN}}(q,z|g^2)
  \ =\ \sum_{\Lambda}q^{-\frac{1}{2}\frac{g^2}{1+g^2}C_\Lambda}
       \,\psi_\Lambda^{\text{WZW}}(q)\,\chi_\Lambda(z)\ \ ,
\end{align}
  where $\psi_\Lambda^{\text{WZW}}(q)$ denotes the branching functions
  at zero coupling. Under the present circumstances the decomposition
  of the WZW spectrum \eqref{eq:WZWSpectrum} into representations of
  $\text{OSP}(4|2)$ can actually be performed explicitly
 \cite{Mitev:2008yt}, resulting in
\begin{equation}
  \begin{split}
  \psi_{[j_1,j_2,j_3]}^{\text{WZW}}(q) &\  = \
\frac{1}{\eta(q)^4}\sum_{n,m=0}^{\infty}(-1)^{n+m}
q^{\frac{m}{2}(m+4j_1+2n+1)+j_1+\frac{n}{2}-\frac{1}{8}}
\\[2mm]& \hspace*{1cm}\times\
\Bigl[q^{(j_2-\frac{n}{2})^2}-q^{(j_2+\frac{n}{2}+1)^2}\Bigr]
\Bigl[q^{(j_3-\frac{n}{2})^2}-q^{(j_3+\frac{n}{2}+1)^2}\Bigr]\ \ .
  \end{split}
\end{equation}
  We recognize from eq.~\eqref{eq:Interpolation} that the deformed
  branching functions have a very simple dependence on the coupling
  $g$.

  Let us discuss the consequences of formula \eqref{eq:Interpolation}
  in more detail, see also figure \ref{fig:Spectrum}. At
  zero coupling, the spectrum is characterized by the following
  features: All states have either integer or half-integer energy and
  at each energy level there is only a finite number of states. As
  mentioned above, these states are accounted for by the two affine
  $\widehat{\text{OSP}}(4|2)_1$ representations built on top of the
  vacuum (with $h=0$) and the fundamental representation (with
  $h=1/2$), respectively. Once the deformation is switched on, the
  affine symmetry is broken and the states will receive an anomalous
  dimension depending on their transformation behavior under global
  $\text{OSP}(4|2)$ transformations (the zero-modes of the current
  algebra). In particular, multiplets belonging to a representation
  with vanishing Casimir do not receive any correction. These are all
  protected BPS representations.\footnote{It
    should be noted, however, that there are short/BPS representations for
    $\text{OSP}(4|2)$ which are not protected in this sense.} This applies
  in particular to the adjoint representation and ensures that the
  currents stay at conformal dimension $h=1$.

  At intermediate coupling the spectrum is very complicated,
  exhibiting almost no sign of an underlying organizing
  principle, except for the preserved global $\text{G}$ and the
  Virasoro symmetry. However, at infinite coupling we again recover a
  special situation. The energy of a multiplet $\Lambda$ is shifted by
  $-C_\Lambda/2$ in this case. It can be shown that despite this shift
  all conformal dimensions remain non-negative. Even more surprising,
  the spectrum is very regular again, exhibiting an integer level
  spacing (as opposed to the half-integer spacing at
  $g=0$). Nevertheless the spectrum now has entirely different
  characteristics than at zero coupling. Indeed, at infinite coupling
  we find an infinite number of states on each energy level, see again
  figure \ref{fig:Spectrum}.

\subsection{\label{sc:SphereSpectrum}Identification with the large
  volume spectrum of a supersphere  \texorpdfstring{$\sigma$}{sigma}-model}

\begin{figure}
\begin{center}
\begin{pgfpicture}{-1cm}{0cm}{10cm}{5cm}
  \pgfsetendarrow{\pgfarrowtriangle{3pt}}
  \pgfsetlinewidth{1.2pt}
  \pgfline{\pgfxy(-1,3.5)}{\pgfxy(3.85,3.5)}
  \pgfputat{\pgfxy(7.7,5)}{\pgfbox[center,center]{$\text{S}^{3|2}$ supersphere $\mathbf{\sigma}$-model}}
  \pgfputat{\pgfxy(4.5,3.5)}{\pgfbox[center,center]{$R$}}
  \pgfputat{\pgfxy(3.5,4)}{\pgfbox[center,center]{Large volume}}
  \pgfputat{\pgfxy(.5,4)}{\pgfbox[center,center]{Quantum regime}}
  {\pgfputat{\pgfxy(2,5)}{\pgfbox[center,center]{\footnotesize geometric}}}
  {\pgfputat{\pgfxy(7.7,3.5)}{\pgfbox[center,center]{$Z_\sigma(q,z|R)$}}}
  \pgfline{\pgfxy(0,1.5)}{\pgfxy(3.85,1.5)}
  \pgfputat{\pgfxy(4.5,1.5)}{\pgfbox[center,center]{$g^2$}}
  \pgfputat{\pgfxy(7.7,0)}{\pgfbox[center,center]{$\text{OSP}(4|2)$ Gross-Neveu model}}
  {\pgfputat{\pgfxy(7.7,1.5)}{\pgfbox[center,center]{$Z_{\text{GN}}(q,z|g^2)$}}}
  \pgfputat{\pgfxy(3.5,1)}{\pgfbox[center,center]{Strong coupling}}
  \pgfputat{\pgfxy(.5,1)}{\pgfbox[center,center]{Weak coupling}}
  {\pgfputat{\pgfxy(2,.2)}{\pgfbox[center,center]{\footnotesize non-geometric}}}
  {\pgfputat{\pgfxy(2,-.2)}{\pgfbox[center,center]{\footnotesize(with potential)}}}
  \pgfclearendarrow
  \pgfsetlinewidth{.4pt}
  \pgfsetdash{{3pt}{3pt}}{0pt}
  {\color{black}\pgfline{\pgfxy(4,3.5)}{\pgfxy(4,1.5)}}
  {\color{black}\pgfline{\pgfxy(0,3.5)}{\pgfxy(0,1.5)}}
  {\pgfputat{\pgfxy(2,2.5)}{\pgfbox[center,center]{$R^2=1+g^2$}}}
  {\color{blue}\pgfcircle[fill]{\pgfxy(0,1.5)}{3pt}}
  {\red\pgfcircle[fill]{\pgfxy(4,1.5)}{3pt}}
  {\color{blue}\pgfcircle[fill]{\pgfxy(4,3.5)}{3pt}}
  {\red\pgfcircle[fill]{\pgfxy(-1,3.5)}{3pt}}
  {\violet\pgfcircle[fill]{\pgfxy(0,3.5)}{3pt}}
  {\pgfsetdash{{2cm}{0pt}}{0pt}\color{black}\pgfline{\pgfxy(7.68,1.7)}{\pgfxy(7.68,3.2)}\pgfline{\pgfxy(7.72,1.7)}{\pgfxy(7.72,3.2)}}
  \pgfputat{\pgfxy(4,5)}{\pgfbox[center,center]{\pgfimage[height=1cm]{WireSphere}}}  
  \pgfputat{\pgfxy(0,5)}{\pgfbox[center,center]{\pgfimage[height=.4cm]{WireSphere}}}  
  \pgfputat{\pgfxy(0,0)}{\pgfbox[center,center]{\pgfimage[height=1cm]{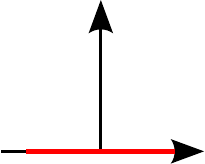}}}  
  \pgfputat{\pgfxy(4,0)}{\pgfbox[center,center]{\pgfimage[height=1cm]{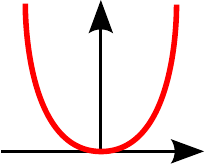}}}  
\end{pgfpicture}
\end{center}
  \caption{\label{fig:Duality}(Color online) The duality between the
    $\text{OSP}(4|2)$ Gross-Neveu model and the $\text{S}^{3|2}$
    supersphere $\sigma$-model in pictures.}
\end{figure}

  We now wish to argue that the spectrum of the Gross-Neveu model
  discussed in the previous section coincides with the large volume
  partition function of the $\sigma$-model on the supersphere
  $\text{S}^{3|2}$ when we send the coupling $g$ to infinity. At infinite
  volume the partition function is easy to write down since the fields
  $\vec{X}$ become free. The most general field is obtained by
  considering the normal ordered products
  $\prod\partial^{n_i}\bartial^{m_i}X^{a_i}$ of the fields $X^a$ and
  their derivatives and the energy (scaling dimension) of such a field
  is just given by the number of derivatives.

  We assume Neumann boundary conditions, i.e.\ a freely moving open
  string. In this case we are only left with one type of
  derivatives. In close analogy to the harmonic analysis on $\text{S}^{3|2}$
  we can write down the open string partition function
\begin{multline}
  \label{eq:ZSSS}
  Z_{\text{S}^{3|2}}(q,z|R=\infty)\\[2mm]
  \ =\ \lim_{t\to1}\prod_{n=1}^\infty\frac{(1-t^2q^n)(1+tq^nz_3)(1+tq^n/z_3)}{(1-tq^nz_1z_2)(1-tq^nz_1/z_2)(1-tq^nz_2/z_1)(1-tq^n/z_1z_2)}\
  \ .
\end{multline}
  The only difference compared to eq.~\eqref{eq:ZSS1} are the
  additional terms involving powers $q^n$. These correspond to counting
  derivatives $\partial^n\vec{X}$ instead of plain coordinates
  $\vec{X}$. Since the constraint $\vec{X}^2=R^2$ also leads to
  constraints on derivatives of $\vec{X}$ also the first term in the
  numerator had to be extended to an infinite product.

  Even though it is by no means obvious, the decomposition of the
  partition function \eqref{eq:ZSSS} into irreducible characters of
  $\text{OSP}(4|2)$ precisely agrees with the limit $g\to\infty$ of the
  expression \eqref{eq:Interpolation}
  \cite{Mitev:2008yt}. This suggests that the moduli spaces of the two
  theories indeed overlap -- and employing their common symmetry --
  actually coincide, see Fig.~\ref{fig:Duality} for an
  illustration. Complementary calculations based on either lattice
  models \cite{Candu:2008yw}, background field methods or a
  cohomological reduction \cite{Candu:2010yg} confirm this picture and
  predict that the couplings should in fact be related as
  $R^2=1+g^2$.

\section{\label{sc:QuasiAbelian}Quasi-abelian perturbation theory}

  In this section we will explain the origin of the formulas
  \eqref{eq:AnomalousDimension2} and \eqref{eq:Interpolation} for the
  distinguished boundary spectra of deformed WZW models. For
  simplicity we restrict our presentation to the case of the
  $\text{G}$-preserving deformation discussed in section
  \ref{sc:WZWDef2}. The $\text{G}\times\text{G}$ preserving deformation can be
  discussed along similar lines but the analysis is slightly more
  complicated due to the presence of the additional field
  $\phi_{\mu\nu}(z,\bz)$ \cite{Quella:2007sg}. Our presentation here
  will be based on a simple analogy with the free boson.

\subsection{Anomalous dimensions}

  In order to determine the anomalous dimensions of boundary operators
  $\psi_\mu(x)$ we need to calculate the two-point correlation
  functions
\begin{align}
  \bigl\langle\psi_\mu(x)\psi_\nu(y)\bigr\rangle_\lambda
  \ =\ \frac{C_{\mu\nu}(\lambda)}{(x-y)^{\Delta_\mu(\lambda)+\Delta_\nu(\lambda)}}\
  \ .
\end{align}
  It should be emphasized that $\Delta(\lambda)$ will in general have
  a diagonal contribution $h(\lambda)$ and a nilpotent part
  $\delta(\lambda)$.  The two-point function is determined
  perturbatively through the formula
\begin{multline}
  \label{eq:Perturbation}
  \bigl\langle\psi_\mu(x)\psi_\nu(y)\bigr\rangle_\lambda
  \ =\ \Bigl\langle\psi_\mu(x)\psi_\nu(y) e^{-\lambda\cS_{\text{def}}}\Bigr\rangle_{\text{WZW}}\\[2mm]
  \ =\
  \sum_{n=0}^\infty\frac{(-\lambda)^n}{n!}\biggl\langle\psi_\mu(x)\psi_\nu(w)\prod_{i=1}^n\int\!d^2w_i\,\bigl\langle
  J(w_i),\bJ(\bw_i)\bigr\rangle\biggr\rangle_{\text{WZW}}\ \ .
\end{multline}
  As usual, the expression in the second line needs regularization and
  renormalization to be well defined. We wish to avoid entering the
  technical details here but rather to make an analogy with the free
  boson.

  In general, it is impossible to make precise statements about
  correlations function beyond a few orders in perturbation theory. The
  same applies here with regard to the structure functions
  $C_{\mu\nu}(\lambda)$ and the nilpotent part of $\Delta(\lambda)$. On
  the other hand the diagonal part $h(\lambda)$ of $\Delta(\lambda)$
  is accessible since it is a $\text{G}$-invariant. In other words, for the
  calculation of $h(\lambda)$ all terms involving structure constants
  can be dropped, following the reasoning of section
  \ref{sc:ConformalInvariance} (cf.\ \eqref{eq:WZWOPE} with
  \eqref{eq:FreeOPE}). The combinatorics (and also the regularization
  and renormalization) hence effectively reduce to those of the radius
  perturbation of a multi-component free boson. For the latter the
  conformal dimensions are known explicitly for all values of the
  radius even without making use of perturbation theory. We are thus
  in the comfortable position to avoid any tedious combinatorics.

  Let us now fix the self-dual radius $R_0$ as a reference radius. The
  free boson at all other radii $R=R_0\sqrt{1+\lambda}$ can be considered
  as a current-current deformation of the type $\langle J,\bJ\rangle$
  of the reference theory. Looking at the concrete
  expression \eqref{eq:BosonPF} for the partition function we find the
  anomalous dimension of a field with momentum quantum number $w$,
\begin{align}
  \delta h_w(\lambda)
  \ =\ \biggl[\frac{1}{1+\lambda}-1\biggr]\frac{w^2}{2R_0^2}
  \ =\ -\frac{\lambda}{1+\lambda}\,\frac{w^2}{2R_0^2}\ \ .
\end{align}
  The last term should be interpreted as the operator $J_0^2$ acting
  on a boundary field. In the WZW context this would give rise to the
  quadratic Casimir $C_\Lambda$. Taking into account the different
  normalization of the currents and the Lagrangians we finally end up
  with eqs.~\eqref{eq:AnomalousDimension2} and
  \eqref{eq:Interpolation}. A more detailed discussion of these issues
  can be found in \cite{Quella:2007sg,Mitev:2008yt}.

\section{Conclusions and Outlook}

  We presented a few of the main peculiarities of conformal
  $\sigma$-models on supergroups $\text{G}$ and supercosets $\text{G/H}$. In these
  models conformal invariance is closely tied to the condition that
  $\text{G}$ has vanishing Killing form, rendering the model
  quasi-abelian. We discussed the harmonic analysis on $\text{G}$ and
  $\text{G/H}$
  in order to get a hold on the large volume regime of the
  $\sigma$-models. In particular, the emergence of non-chiral
  indecomposable representations was pointed out as a consequence of
  supergeometry. Afterwards we introduced supergroup WZW and their
  deformations. As a concrete example we presented the $\text{OSP}(4|2)$
  Gross-Neveu model. Based on an exact perturbative expression for a
  boundary spectrum we finally argued for a duality between the
  non-geometric Gross-Neveu model and the $\text{S}^{3|2}$ supersphere
  $\sigma$-model. Throughout the paper we used analogies with the free
  boson to motivate observations and part of the formulas.

  The content of this note can be extended into several directions.
  A more complicated application of supercoset $\sigma$-models
  concerns projective superspaces (cf.\ table \ref{tab:CondMat})
  which, in contrast to superspheres, admit a topological term. In
  that case an additional idea, namely cohomological reduction to a
  symplectic fermion theory, has to be employed in order to derive
  exact boundary partition functions
  \cite{Candu:2009ep,Candu:2010yg}. A different direction concerns the
  study of the current algebras associated with deformed WZW models
  \cite{Ashok:2009xx,Benichou:2010rk,Konechny:2010nq}. Away from the
  WZW point, there is still a current algebra but it becomes
  non-chiral and current three-point functions start exhibiting
  logarithmic singularities. One may assume that a better
  understanding of the interplay between this local current algebra
  and additional non-local conserved charges will provide the key to
  an exact solution of the models beyond the results that have been
  presented here.

\subsubsection*{Acknowledgments}

  TQ is very grateful for the opportunity to present this work at the
  Lorentz Workshop ``The Interface of Integrability and Quantization''
  and for the stimulating discussions with the other
  participants. The research of TQ was partially funded by a Marie
  Curie Intra-European Fellowship, contract number
  MEIF-CT-2007-041765. We furthermore acknowledge partial support from
  the EU Research Training Network {\it Superstring theory},
  MRTN-CT-2004-512194 and from {\it ForcesUniverse},
  MRTN-CT-2004-005104.


\begin{thebibliography}{10}
\expandafter\ifx\csname url\endcsname\relax
  \def\url#1{\texttt{#1}}\fi
\expandafter\ifx\csname urlprefix\endcsname\relax\def\urlprefix{URL }\fi
\expandafter\ifx\csname href\endcsname\relax
  \def\href#1#2{#2} \def\path#1{#1}\fi

\bibitem{Efetov1983:MR708812}
K.~B. Efetov, Supersymmetry and theory of disordered metals, Adv. Phys. 32
  (1983) 53--127.

\bibitem{Parisi:1979ka}
G.~Parisi, N.~Sourlas, Random magnetic fields, supersymmetry and negative
  dimensions, Phys. Rev. Lett. 43 (1979) 744.

\bibitem{Parisi:1982ud}
G.~Parisi, N.~Sourlas, Supersymmetric field theories and stochastic
  differential equations, Nucl. Phys. B206 (1982) 321.

\bibitem{Read:2001pz}
N.~Read, H.~Saleur, Exact spectra of conformal supersymmetric nonlinear sigma
  models in two dimensions, Nucl. Phys. B613 (2001) 409.
\newblock \href {http://arxiv.org/abs/hep-th/0106124}
  {\path{arXiv:hep-th/0106124}}.

\bibitem{Candu:2009pj}
C.~Candu, J.~L. Jacobsen, N.~Read, H.~Saleur, Universality classes of polymer
  melts and conformal sigma models, J. Phys. A43~(14) (2010) 142001.
\newblock \href {http://arxiv.org/abs/0908.1081}
{\path{arXiv:0908.1081}}.

\bibitem{Kagan:2005wt}
D.~Kagan, C.~A.~S. Young, Conformal sigma-models on supercoset targets, Nucl.
  Phys. B745 (2006) 109--122.
\newblock \href {http://arxiv.org/abs/hep-th/0512250}
  {\path{arXiv:hep-th/0512250}}.

\bibitem{Candu:2010yg}
C.~Candu, T.~Creutzig, V.~Mitev, V.~Schomerus, Cohomological reduction of sigma
  models, JHEP 05 (2010) 047.
\newblock \href {http://arxiv.org/abs/1001.1344}
{\path{arXiv:1001.1344}}.

\bibitem{Bershadsky:1999hk}
M.~Bershadsky, S.~Zhukov, A.~Vaintrob, {$PSL(n|n)$} sigma model as a conformal
  field theory, Nucl. Phys. B559 (1999) 205--234.
\newblock \href {http://arxiv.org/abs/hep-th/9902180}
  {\path{arXiv:hep-th/9902180}}.

\bibitem{Kac:1977em}
V.~G. Kac, Lie superalgebras, Adv. Math. 26 (1977) 8--96.

\bibitem{Quella:2007hr}
T.~Quella, V.~Schomerus, Free fermion resolution of supergroup {WZNW} models,
  JHEP 09 (2007) 085.
\newblock \href {http://arxiv.org/abs/0706.0744} {\path{arXiv:0706.0744}}.

\bibitem{Schomerus:2005bf}
V.~Schomerus, H.~Saleur, The {$GL(1|1)$} {WZW} model: {From} supergeometry to
  logarithmic {CFT}, Nucl. Phys. B734 (2006) 221--245.
\newblock \href {http://arxiv.org/abs/hep-th/0510032}
  {\path{arXiv:hep-th/0510032}}.

\bibitem{Mitev:2008yt}
V.~Mitev, T.~Quella, V.~Schomerus, Principal chiral model on superspheres, JHEP
  11 (2008) 086.
\newblock \href {http://arxiv.org/abs/0809.1046} {\path{arXiv:0809.1046}}.

\bibitem{Quella:2007sg}
T.~Quella, V.~Schomerus, T.~Creutzig, Boundary spectra in superspace sigma
  models, JHEP 10 (2008) 024.
\newblock \href {http://arxiv.org/abs/0712.3549} {\path{arXiv:0712.3549}}.

\bibitem{Berkovits:1999im}
N.~Berkovits, C.~Vafa, E.~Witten, Conformal field theory of {AdS} background
  with {Ramond-Ramond} flux, JHEP 03 (1999) 018.
\newblock \href {http://arxiv.org/abs/hep-th/9902098}
  {\path{arXiv:hep-th/9902098}}.

\bibitem{Lesage:2002ch}
F.~Lesage, P.~Mathieu, J.~Rasmussen, H.~Saleur, The
  {$\widehat{su}(2)_{-\frac{1}{2}}$} {WZW} model and the {$\beta\gamma$}
  system, Nucl. Phys. B647 (2002) 363--403.
\newblock \href {http://arxiv.org/abs/hep-th/0207201}
  {\path{arXiv:hep-th/0207201}}.

\bibitem{Lesage:2003kn}
F.~Lesage, P.~Mathieu, J.~Rasmussen, H.~Saleur, Logarithmic lift of the
  {$\widehat{su}(2)_{-\frac{1}{2}}$} model, Nucl. Phys. B686 (2004) 313.
\newblock \href {http://arxiv.org/abs/hep-th/0311039}
  {\path{arXiv:hep-th/0311039}}.

\bibitem{Candu:2008yw}
C.~Candu, H.~Saleur, A lattice approach to the conformal {$OSp(2S+2|2S)$}
  supercoset sigma model. {Part II:} {The} boundary spectrum, Nucl. Phys. B808
  (2009) 487--524.
\newblock \href {http://arxiv.org/abs/0801.0444} {\path{arXiv:0801.0444}}.

\bibitem{Candu:2009ep}
C.~Candu, V.~Mitev, T.~Quella, H.~Saleur, V.~Schomerus, The sigma model on
  complex projective superspaces, JHEP 02 (2010) 015.
\newblock \href {http://arxiv.org/abs/0908.0878} {\path{arXiv:0908.0878}}.

\bibitem{Ashok:2009xx}
S.~K. Ashok, R.~Benichou, J.~Troost, Conformal current algebra in two
  dimensions, JHEP 06 (2009) 017.
\newblock \href {http://arxiv.org/abs/0903.4277} {\path{arXiv:0903.4277}}.

\bibitem{Benichou:2010rk}
R.~Benichou, J.~Troost, The conformal current algebra on supergroups with
  applications to the spectrum and integrability, JHEP 04 (2010) 121.
\newblock \href {http://arxiv.org/abs/1002.3712} {\path{arXiv:1002.3712}}.

\bibitem{Konechny:2010nq}
A.~Konechny, T.~Quella, {Non-chiral current algebras for deformed supergroup
  WZW models}, JHEP 1103 (2011) 124.
\newblock \href {http://arxiv.org/abs/1011.4813} {\path{arXiv:1011.4813}}.

\end{thebibliography}

\def\cprime{$'$}

\end{document}